\documentclass[twocolumn]{aastex631} %

\usepackage{CJK}
\usepackage{amsmath}
\usepackage{xcolor} 
\definecolor{bg-green}{rgb}{0.8588,0.9333,0.8666}
\definecolor{Q-color}{rgb}{0.5,0.1,0.5}

\shorttitle{Mass Loading of AGN-Driven Outflows}
\shortauthors{Qiu et al.}

\begin{document}

\begin{CJK*}{UTF8}{gbsn}

\title{On the Mass Loading of AGN-Driven Outflows in Elliptical Galaxies and Clusters}

\correspondingauthor{Yu Qiu (邱宇)}
\email{yuqiu@pku.edu.cn}

\author[0000-0002-6164-8463]{Yu Qiu (邱宇)}
\affiliation{Kavli Institute for Astronomy and Astrophysics, Peking University, 5 Yiheyuan Road, Haidian District, Beijing, 100871, PRC}

\author[0000-0002-2622-2627]{Brian R. McNamara}
\affiliation{Department of Physics and Astronomy, University of Waterloo, 200 University Avenue West, Waterloo, ON, N2L 3G1, Canada}
\affiliation{Waterloo Center for Astrophysics, University of Waterloo, 200 University Avenue West, Waterloo, ON, N2L 3G1, Canada}
\affiliation{Perimeter Institute for Theoretical Physics, Waterloo, ON, N2L 2Y5, Canada}

\author[0000-0002-7835-7814]{Tamara Bogdanovi\'c}
\affiliation{Center for Relativistic Astrophysics, School of Physics, Georgia Institute of Technology, 837 State Street, Atlanta, GA 30332, USA}

\author[0000-0001-9840-4959]{Kohei Inayoshi}
\affiliation{Kavli Institute for Astronomy and Astrophysics, Peking University, 5 Yiheyuan Road, Haidian District, Beijing, 100871, PRC}

\author[0000-0001-6947-5846]{Luis C. Ho}
\affiliation{Kavli Institute for Astronomy and Astrophysics, Peking University, 5 Yiheyuan Road, Haidian District, Beijing, 100871, PRC}
\affiliation{Department of Astronomy, School of Physics, Peking University, 5 Yiheyuan Road, Haidian District, Beijing 100871, PRC}

\begin{abstract}
Outflows driven by active galactic nuclei (AGN) are an important channel for accreting supermassive black holes (SMBHs) to interact with their host galaxies and clusters. Properties of the outflows are however poorly constrained due to the lack of kinetically resolved data of the hot plasma that permeates the circumgalactic and intracluster space. In this work, we use a single parameter, outflow-to-accretion mass-loading factor $m=\dot{M}_{\rm jet}/\dot{M}_{\rm BH}$, to characterize the outflows that mediate the interaction between SMBHs and their hosts. By modeling both M87 and Perseus, and comparing the simulated thermal profiles with the X-ray observations of these two systems, we demonstrate that $m$ can be constrained between $200-500$. This parameter corresponds to a bulk flow speed between $4,000-7,000\,{\rm km\,s}^{-1}$ at around 1\,kpc, and a thermalized outflow temperature between $10^{8.7}-10^{9}\,{\rm K}$. Our results indicate that the dominant outflow speeds in giant elliptical galaxies and clusters are much lower than in the close vicinity of the SMBH, signaling an efficient coupling with and deceleration by the surrounding medium on length scales below 1\,kpc. Consequently, AGNs may be efficient at launching outflows $\sim10$ times more massive than previously uncovered by measurements of cold, obscuring material. We also examine the mass and velocity distribution of the cold gas, which ultimately forms a rotationally supported disk in simulated clusters. The rarity of such disks in observations indicates that further investigations are needed to understand the evolution of the cold gas after it forms.
\end{abstract}

\keywords{Cooling flows(2028), Intracluster medium(858), Active galactic nuclei(16), Jets(870), Virgo Cluster(1772), Perseus Cluster(1214)}

\section{Introduction} \label{sec:intro}
Studies of supermassive black holes (SMBHs) and their host galaxies have revealed a correlation between black hole mass and galaxy properties such as bulge mass, luminosity, and velocity dispersion \citep{Magorrian1998, Ferrarese2000, Gebhardt2000}, suggesting a coevolution scenario \citep{Kormendy2013}. The most promising mechanism that mediates their coevolution is energy released by accreting SMBHs, also known as active galactic nuclei (AGN) feedback \citep{Fabian2012}. The main modes of AGN feedback include radiation \citep{Silk1998}, accretion disk winds \citep{Murray1995}, and relativistic jets \citep[see][for a review]{Blandford2019}. In particular, the jet mode is more common in elliptical galaxies, groups, and clusters, due to their relatively low SMBH accretion rates \citep{McNamara2012}. Radio-emitting relativistic jets from AGNs can drive large scale outflows and inflate cavities in the X-ray-emitting circumgalactic and intracluster medium, thus doing mechanical work to heat the hot plasma and prevent further star formation {\citep[e.g.,][]{Birzan2004, McNamara2005}}. This feedback mode is therefore also referred to as mechanical or radio-mode feedback.

However, the mechanism that couples jets with the surrounding medium is largely unknown. At launch, jets are likely dominated by relativistic electron-positron `pair' plasma at both low and high radio luminosities \citep{Reynolds1996a, Wardle1998}. The dissipation of FR I jets \citep{Fanaroff1974}, as well as X-ray cavities filled with radio plasma \citep{Birzan2004}, indicates that the jets are decelerated by baryons along their paths. This process involves interaction between the jet material and the stellar wind or gas clouds surrounding the central AGN {\citep[e.g.,][]{Komissarov1994, Bowman1996, Hubbard2006, Wagner2011, Bosch-Ramon2012, Wagner2012, Morganti2013, Walg2013, Perucho2014, Cielo2014, Mukherjee2016, Mukherjee2018, Cielo2018, Angles-Castillo2021}}. The resulting outflows are therefore mass-loaded, and provide a channel for SMBHs to interact with their hosts.

Gas accretion and outflows driven by the central SMBH constitute a natural regulating cycle that tightly binds the SMBH with the host. Many recent studies of elliptical galaxies and clusters have simulated the feedback cycles by using physically motivated models of jets. Due to the large dynamical range from the vicinity of the SMBH to galaxy cluster outskirts, jet feedback is routinely implemented as the mass-loaded baryonic outflows rather than the original relativistic pair plasma. The outflow speed, however, are often selected arbitrarily $\gtrsim10,000\,{\rm km\,s}^{-1}$, such as works of \citet{Omma2004, Sternberg2007, Dubois2010a, Gaspari2012a, Choi2012, Li2014}; and \citet{Wang2018}.\footnote{Note that in some works, AGN `wind' rather than jet is considered the main mechanism in driving outflows. We discuss this scenario in Section~\ref{sec:implication}.} This is on par with the maximum velocities measured by the broad/narrow quasar absorption line widths/shifts \citep{Crenshaw2003, Tombesi2010}. As we will explain in Section~\ref{sec:method}, this speed choice implies an outflow-to-accretion mass-loading factor, $m\equiv\dot{M}_{\rm jet}/\dot{M}_{\rm BH}<100$, {assuming 10\% of the accreted rest mass energy is channeled into the outflows}. However, the maximum velocity component does not necessarily represent the bulk of the flow, and further investigation on the mass-loading factor of AGN-driven outflows is needed.

Observationally, the mass-loading factor $m$ is poorly constrained. Measurements of AGN-driven outflows depend heavily on the absorption of an obscuring material or the emission of the surrounding cold gas clouds, which only cover a portion of the multiphase outflow. Energetically, however, such observed outflows do not seem to contribute to the heating of the hot plasma. We have shown in our previous works that outflows with initial temperature $T\lesssim 10^7\,{\rm K}$ will cool quickly to form cold gas filaments, and constitute a channel for positive AGN feedback that ultimately elevates the formation of cold gas \citep{Qiu2019, Qiu2018, Qiu2020}. Therefore, only X-ray-emitting outflows with higher temperature, or equivalently, high speed outflows that quickly thermalize, can create a negative feedback channel to reduce cold gas production. {From a theoretical point of view, a smaller $m$ indicates less interaction between the jet and the baryons and less efficient heating of the medium in the vicinity of the central AGN. Beyond a few kiloparsecs, however, a smaller $m$ leads to a hotter outflow that may be more capable at heating the ICM. Our simulations are therefore designed to test this hypothesis.} Since the outflows are multiphase, observational and theoretical studies that focus on a single component will likely recover only part of the whole distribution. Nevertheless, isolating the dominant component is crucial in uncovering the underlying link between SMBHs and their hosts. 

In this work, our aim is to use 3D AGN jet feedback simulations to explore how varying mass-loading factors impact the evolution of the hot plasma in representative elliptical galaxies and clusters. To better constrain this parameter and reveal the dominant outflow component associated with negative AGN feedback, we also compare our results with the X-ray observations of the hot plasma. We explain our simulation setup and feedback modeling in Section~\ref{sec:method}, present the results in Section~\ref{sec:results}, discuss the implications in Section~\ref{sec:discussion}, and conclude in Section~\ref{sec:conclusion}. 

\section{Methodology} \label{sec:method}

\begin{deluxetable*}{lccclrrccc}[t!]
\tablecaption{Simulation Parameters}
\tablewidth{0pt}
\tablehead{
\colhead{Simulation} & \colhead{Res.} & \colhead{$\varepsilon_{\rm acc}$} & \colhead{$v_{\rm out}$} & \colhead{$T_{\rm out}$} & \colhead{$m^\dagger$}  & \colhead{$\langle\dot{M}_{\rm BH}\rangle^*$} & \colhead{$f_{\rm duty}^\ddagger$} & \colhead{$f_{\rm R43}^\star$} \\
\colhead{ID} & \colhead{(kpc)} & \colhead{} & \colhead{($10^3\,{\rm km\,s}^{-1}$)} & \colhead{(K)} & \colhead{}
 & \colhead{($M_\sun\,{\rm yr}^{-1}$)}  & \colhead{}  & \colhead{}
}
\startdata
P8.3		& 0.49	& $10^{-2}$	& 2.97	& $10^{8.33}$	& 1017 	& 2.14	& -		& -	\\
P8.7		& 0.49	& $10^{-2}$	& 4.40	& $10^{8.67}$	& 465 	& 0.96	& -		& -	\\
P8.7hr	& 0.24	& $10^{-2}$	& 4.40	& $10^{8.67}$	& 465 	& 1.14	& -		& -	\\
P9.0		& 0.49	& $10^{-2}$	& 6.43	& $10^{9.0}$	& 217 	& 0.40	& -		& -	\\
P10		& 0.49	& $10^{-2}$	& 20.3	& $10^{10}$	& 21.7 	& 0.58	& -		& -	\\
M8.3		& 0.49	& $10^{-2}$	& 2.97	& $10^{8.33}$	& 1017 	& 0.0044	& 0.24	& 0.066	\\
M8.7he	& 0.49	& $10^{-1}$	& 4.40	& $10^{8.67}$	& 465 	& 0.0051	& -		& -	\\
M8.7		& 0.49	& $10^{-2}$	& 4.40	& $10^{8.67}$	& 465 	& 0.0036	& 0.35	& 0.032	\\
M8.7le	& 0.49	& $10^{-3}$	& 4.40	& $10^{8.67}$	& 465 	& 0.0033	& -		& -	\\
M9.0		& 0.49	& $10^{-2}$	& 6.43	& $10^{9.0}$	& 217 	& 0.0041	& 0.51	& 0.055	\\
\enddata
\label{tab:params}
\tablecomments{Simulation IDs start with a letter denoting the modeled system, (P)erseus or (M)87, followed by a number corresponding to the base-10 logarithm of the characteristic outflow temperature $T_{\rm out}$. Appended letters represent: (hr) high resolution, (le) low accretion efficiency, and (he) high accretion efficiency. The outflow speed $v_{\rm out}$, outflow temperature $T_{\rm out}$, and mass-loading factor $m$ are interdependent based on the model laid out in Equation~\ref{eqn:model}, so we use $T_{\rm out}$ for reference throughout the paper. $^\dagger$Values of the mass-loading factor $m$ are based on the assumption that the feedback efficiency $\eta=0.1$. $^*$In Perseus simulations, $\langle\dot{M}_{\rm BH}\rangle$ is averaged over the evolution after 1\,Gyr, while in M87 simulations, the entire 2\,Gyr. {$^\ddagger$In M87 simulations where AGN jet luminosity has a cyclic behavior, we characterize the duty cycle as the fraction of time when the AGN jet is operating in the high luminosity stage ($L_{\rm J}\gtrsim10^{43}\,{\rm erg\,s}^{-1}$, as described in Appendix~\ref{app:duty}). $^\star$We also infer the radiative output of the AGN at low accretion rates, and calculate the fraction of time AGN radiative luminosity $L_{\rm R}>10^{43}\,{\rm erg\,s}^{-1}$, shown as $f_{\rm R43}$.}}
\end{deluxetable*}

The simulations are performed using the hydrodynamic code \texttt{Enzo} \citep{Bryan2014}, {and is set up similarly to \citet{Qiu2018} on an adaptive mesh refinement grid based on gas density and cooling time, with the highest resolution equal to 0.24\,kpc or 0.49\,kpc. We refer readers to the previous publication for the details of the refinement criteria}, and focus our method description on the physical processes involved in this work. We initialize the simulation domain of $(500\,{\rm kpc})^3$ using observed density and temperature profiles of M87 or Perseus, and provide a static background gravitational potential for the gas to evolve. In the central $r<1\,{\rm kpc}$, we estimate the accretion rate of the SMBH, $\dot{M}_{\rm BH}$, based on the gas properties, and convert it to AGN jet power. The simulated systems are then evolved for $\sim2\,{\rm Gyr}$. In the following subsections, we will describe each process in detail. 

\subsection{Black Hole Accretion}\label{sec:accretion}
{The nominal accretion radius, $r_{\rm a}=1\,{\rm kpc}$, corresponds to the Bondi accretion radius for temperature $T\sim10^6\,{\rm K}$ gas, around a $M_{\rm BH}\sim3.5\times10^9\,M_\sun$ black hole}. To estimate the black hole accretion rate, we consider the accretion of hot, ionized plasma \citep[{modeled as Bondi-Hoyle-Lyttleton accretion;}][]{Bondi1952} and the accretion of cold gas \citep[estimated from its mass and free-fall time;][]{Pizzolato2005}. {Due to resolution limits, the Bondi radius for gas with $T>10^6\,{\rm K}$ is not resolved, but we note that the accretion process is dominated by the cold gas at the peaks of AGN activity \citep[e.g.,][]{Qiu2018}.} In this method, ``cold'' is defined as $T<3\times10^4\,{\rm K}$. The accretion rate is then taken as the sum of the two modes, namely:
\begin{equation}
	\dot{M}_{\rm BH}=\varepsilon_{\rm acc}\left[\frac{M_{\rm cg}}{\tau}+\frac{4\pi G^2\rho_\infty M_{\rm BH}^2}{\left(c_{\rm s}^2+v_{\rm g}^2\right)^{3/2}}\right],
\label{eqn:mdot}	
\end{equation}
where $\varepsilon_{\rm acc}$ is the accretion efficiency, $M_{\rm cg}$ is the amount of cold gas, $\tau = 5\,{\rm Myr}$ is the characteristic free-fall time at 1\,kpc, $\rho_\infty$ is the average density of the gas, $c_{\rm s}$ is the average sound speed calculated using the mass-weighted temperature, and $v_{\rm g}$ is the mass-weighted average velocity of the gas, all calculated within $r_{\rm a}$. {This setup ensures that the SMBH accretion and AGN feedback always operate, with varying accretion rates and luminosities based on the evolution of the core gas properties.} Following \citet{Qiu2018}, $\varepsilon_{\rm acc}$ is taken to be $10^{-2}$ in all Perseus simulations. {Observationally, the gas accretion rate in M87 is found to drop from $\sim 0.1\,M_\sun\,{\rm yr}^{-1}$ at the Bondi radius \citep{Russell2015}, to $\lesssim10^{-3}\,M_\sun\,{\rm yr}^{-1}$ at 21 Schwarzschild radii \citep{Kuo2014}, which also implies an accretion efficiency $\lesssim10^{-2}$.} This parameter is experimented with in the M87 setup and discussed in Appendix~\ref{app:resolution}. {Pairing the estimated SMBH accretion rate of $\lesssim10^{-3}\,M_\sun\,{\rm yr}^{-1}$ with a typical jet power estimate of $>10^{43}\,{\rm erg\,s}^{-1}$ in M87 \citep[e.g.,][]{Forman2017}, the conversion factor from rest mass to jet energy is on the order of 10\% (see the parameter $\eta$ below).} A full description of the simulation parameters is summarized in Table~\ref{tab:params}.

\subsection{Jet Modeling}\label{sec:jet}
After obtaining the SMBH accretion rate, 10\% of the accreted rest mass energy is channeled to the energy of the jet-mode feedback, which is then divided between kinetic and thermal components to account for internal shock heating within $r_{\rm a}$, i.e.,
\begin{equation}
\label{eqn:model}
\begin{split}
L_{\rm J}&=\eta \dot{M}_{\rm BH} c^2 = L_{\rm k}+L_{\rm t},\\
L_{\rm k}&=\frac{1}{2}\dot{M}_{\rm jet}  v_{\rm out}^2,\\
L_{\rm t}&=\frac{\dot{M}_{\rm jet}k_{\rm B}T_{\rm out}}{(\gamma-1)\mu m_p}
\end{split}
\end{equation} 
{where $\eta=0.1$ is the jet feedback efficiency. We fix this value based on the observations of M87 mentioned above. This value is also tested by cosmological simulations such as \citet{Dubois2012}, who find that the conversion factor from accreted rest mass energy to radio-mode feedback needs to be $\sim0.1$ in order for BH densities to be consistent with observational constraints \citep[This is higher than the overall efficiency factor of 0.015 calibrated in][which employs purely thermal feedback]{Booth2009}. We note, however, the magnetically arrested disc model has suggested that the feedback efficiency can exceed 100\% depending on the spin of the BH \citep{Tchekhovskoy2011}.} $\dot{M}_{\rm jet}$ and $v_{\rm out}$ are respectively the mass rate and velocity of the outflowing gas. $L_{\rm k}$ and $L_{\rm t}$ are the kinetic and thermal feedback luminosity. $k_{\rm B}$ is the Boltzmann constant, $\gamma=5/3$ is the adiabatic index for monatomic ideal gas, $\mu=0.6$ is the mean atomic weight for fully ionized plasma, and $m_p$ is the proton mass. {For outflows with $L_{\rm t}<L_{\rm k}$}, kinetic energy can be quickly thermalized through shock heating, we therefore fix $L_{\rm k}\equiv L_{\rm t}$ at launch, and note that for $T_{\rm out}>10^8\,{\rm K}$, the added thermal energy does not radiate away efficiently. \citet{Li2014a} explored the allocation of power between kinetic and thermal components, and find no significant impact on the long term thermal balance of the plasma. {Deviations from this equipartition, e.g., $L_{\rm k}<L_{\rm t}$ means lower $v_{\rm out}$ and higher $T_{\rm out}$ compared with the quoted values in Table~\ref{tab:params}, or vice versa. We note that simulations that resolve a much smaller scale \citep[e.g.,][]{Wagner2012, Cielo2014} are needed\footnote{{For example, simulations in \citet{Cielo2014} show for initially hot ($T\sim10^{10}\,{\rm K}$) jets, embedded in an interstellar medium with idealized conditions, launched at $>100,000\,{\rm km\,s}^{-1}$ speeds, the kinetic energy does not thermalize efficiently.}} in order to fully understand the energy partition below 1\,kpc.} This enforced balance of kinetic and thermal energy implies an interdependent relation among $m$ (see below), $T_{\rm out}$, and $v_{\rm out}$. We use the thermalized outflow temperature, $T_{\rm out}$, to characterize each simulation. Three values are selected between $10^8-10^9\,{\rm K}$ based on previous simulation analysis \citep{Qiu2018, Qiu2020}. An additional Perseus simulation with $T_{\rm out}=10^{10}\,{\rm K}$ is also modeled, which is comparable to other simulation studies that employ fast outflows $>10^4\,{\rm km\,s}^{-1}$. {With $v_{\rm out}\gtrsim3,000\,{\rm km\,s}^{-1}$, we note that in the absence of non-gravitational processes, the outflows have enough kinetic energy to travel beyond 1\,Mpc in the potential well defined below.} The parameter space explored is summarized in Table~\ref{tab:params}.

Note that in order to focus on the mass-loading of jets, in this work we are omitting the radiative feedback that comes into play at high accretion rates, which was studied in detail in \citet{Qiu2018}. This setup is consistent with observations of M87, which has a radiatively inefficient AGN and a low SMBH accretion rate. But as we will see in Section~\ref{sec:results}, the omission of radiative feedback may contribute to higher average accretion rates in more massive systems such as the simulated Perseus cluster.

While in reality the jet-driven outflows contain multiphase gas traveling at varying speeds, the choice to model only one initial speed allows us to isolate the component that dominates in the velocity distribution. Following Equation~\ref{eqn:model}, this also indicates that the mass-loading factor can be expressed as:
\begin{equation}
m\equiv\frac{\dot{M}_{\rm jet}}{\dot{M}_{\rm BH}}=\frac{\eta c^2}{v_{\rm out}^2}.
\end{equation}

In each simulation timestep $\Delta t$, the accreted mass $\dot{M}_{\rm BH}\Delta t$ is removed from the accretion region within $r_{\rm a}$ in proportion to individual cell mass. An additional mass component, $\dot{M}_{\rm jet}\Delta t$, is also removed from this region and equally displaced to two launching planes perpendicular to the $z$-axis. These planes are single-cell in thickness, 2 cell widths ($2\Delta z$) in radius, and offset by $2\Delta z$ from the central SMBH. The distribution of added mass in each gas cell on the planes follows a Gaussian distribution, as a function of its distance to the $z$-axis ($r_{z,i}$), i.e., $\Delta M_i\propto e^{-[r_{z,i}/(2\Delta z)]^2/2}$, with a normalization factor that totals to $\dot{M}_{\rm jet}\Delta t$. The cells in the planes are then loaded with additional momentum and thermal energy, i.e., $\pm \Delta M_i v_{\rm out}$ along the $z$-axis, and $\frac{\Delta M_i k_{\rm B}T_{\rm out}}{(\gamma-1)\mu m_p}$, to complete the jet implementation. 

{We caution, however, unlike the jet implementation in \citet{Qiu2018} that directly accelerates the gas around the central AGN, in the current jet model there may be numerical loss of the kinetic energy \citep[see, e.g.,][]{Bourne2017}. Due to the redistribution of gas mass within the accretion region, the jet momentum is mixed with the surrounding gas. This means that the kinetic energy ($K$) has a difference before and after the mixing:}
\begin{equation}
\begin{split}
\Delta K_{i} &=K_{i,{\rm mix}}-K_{i}-K_{{\rm out}}\\
&=\frac{\left(M_i v_i+\Delta M_i v_{\rm out}\right)^2}{2\left(M_i+\Delta M_i\right)}-\frac{1}{2}M_i v_i^2 - \frac{1}{2}\Delta M_i v_{\rm out}^2\\
&=-\frac{ M_i \Delta M_i}{2 (M_i+\Delta M_i)}\left(v_{\rm out}-v_i\right)^2\leq0.
\end{split}
\end{equation}
{The difference vanishes in the steady state where $v_i = v_{\rm out}$ is achieved. We have experimented with a modified jet model where the kinetic energy loss is compensated for in the thermal form, i.e., we add $-\Delta K_{i}$ as the thermal energy of the gas cell. The difference in the simulated cluster is however negligible in terms of the quantities we examine (i.e., evolutions of AGN luminosity, cold gas mass, and the thermal properties of the ICM), because the outflow properties quickly reach a steady state in each AGN outburst.}

This jet setup is similar to that of \citet{Li2014}, with an update on the treatment of accretion and outflow mass rates, and no presumed jet precession. {While we do not explicitly model the re-orientation of jets \citep[such as the jet precession or BH spin evolution modeled in][]{Gaspari2012a,  Li2014, Li2014a, Cielo2018, Beckmann2019}, the outflows may be scattered by dense gas clouds near the AGN, thus redistributing the feedback energy to a broader angular range \citep[see also][]{Qiu2018}. We will later show that the ICM profiles and cold gas mass in our simulation P10 are similar to studies that model jet precession \citep[e.g.,][]{Gaspari2012a, Li2014, Li2014a}.}

\subsection{Initialization of the hot plasma}\label{sec:galaxy}
In this work we focus our modeling on the evolution of the X-ray emitting plasma as it cools radiatively or is heated by the AGN-driven outflows described above. The cooling rate, as well as the initial gas profiles and background dark matter, stellar mass, and black hole mass of the Perseus cluster is already documented in \citet{Qiu2018}, so in this section we supplement only the gas profiles and background gravity components used for M87 simulations. {For simplicity we do not model the gas self-gravity, star formation, or stellar feedback processes, as we focus mainly on the heating and cooling of the ICM. We note, however, the star formation rate in most central galaxies is around $1-10\%$ of the mass cooling rate inferred from X-ray observations \citep{McDonald2018} and may help consume part of the cold gas.}

The stellar and dark matter density profiles are modeled after stellar dynamics fits from \citet{Romanowsky2001}. In particular, the NFW dark matter profile \citep{Navarro1996} is described by the scale radius $r_{\rm s} = 68\,{\rm kpc}$ and the scale density that satisfies $\rho_{\rm s}r_{\rm s}^3=2.4\times 10^{12}\,M_\sun$, which means the dark matter density:
\begin{equation}
\rho_{\rm DM}(r)=\frac{\rho_{\rm s}r_{\rm s}^3}{r(r+r_{\rm s})^2}.
\end{equation}

For the stellar density that starts to dominate in the inner $\sim10\,{\rm kpc}$, we adopt the profile:
\begin{equation}
\rho_{*}(r)=\frac{(3-\beta)M_*}{4\pi a^3}\left(\frac{a}{r}\right)^\beta\left(\frac{a}{r+a}\right)^{4-\beta},
\end{equation}
where $M_*=5\times 10^{11}\,M_\sun$, $a=4\,{\rm kpc}$, and $\beta=1.2$, which is in agreement with the parametrized model fit in Section~2.1 of \citet{Romanowsky2001}. The dependence on $r$ at large radii as $\propto r^{-4}$ ensures agreement with the empirical luminosity profile found in \citet{DeVaucouleurs1948}. For comparison, $\beta=2$ and $\beta=1$ correspond to profiles proposed in \citet{Jaffe1983} and \citet{Hernquist1990}, respectively. The profile also has a convenient functional form for the enclosed stellar mass:
\begin{equation}
M_{\rm *,enc}(r)=M_* \left(\frac{r}{r+a}\right)^{3-\beta}.
\end{equation}
The last component of the static gravitational potential is the SMBH. While there is still uncertainties in the M87 black hole mass based on stellar or gas dynamics measurements \citep{Gebhardt2011, Walsh2013}, it is subdominant on scales $r\gtrsim1\,{\rm kpc}$. We nevertheless choose $M_{\rm BH}=3.5\times10^{9}\,M_\sun$ for completeness. The choice of the black hole mass has no implication for jet feedback power other than scaling the Bondi accretion rate, as shown in Equation~\ref{eqn:mdot}.

\begin{figure*}[t!]
\centering
\includegraphics[width=180mm]{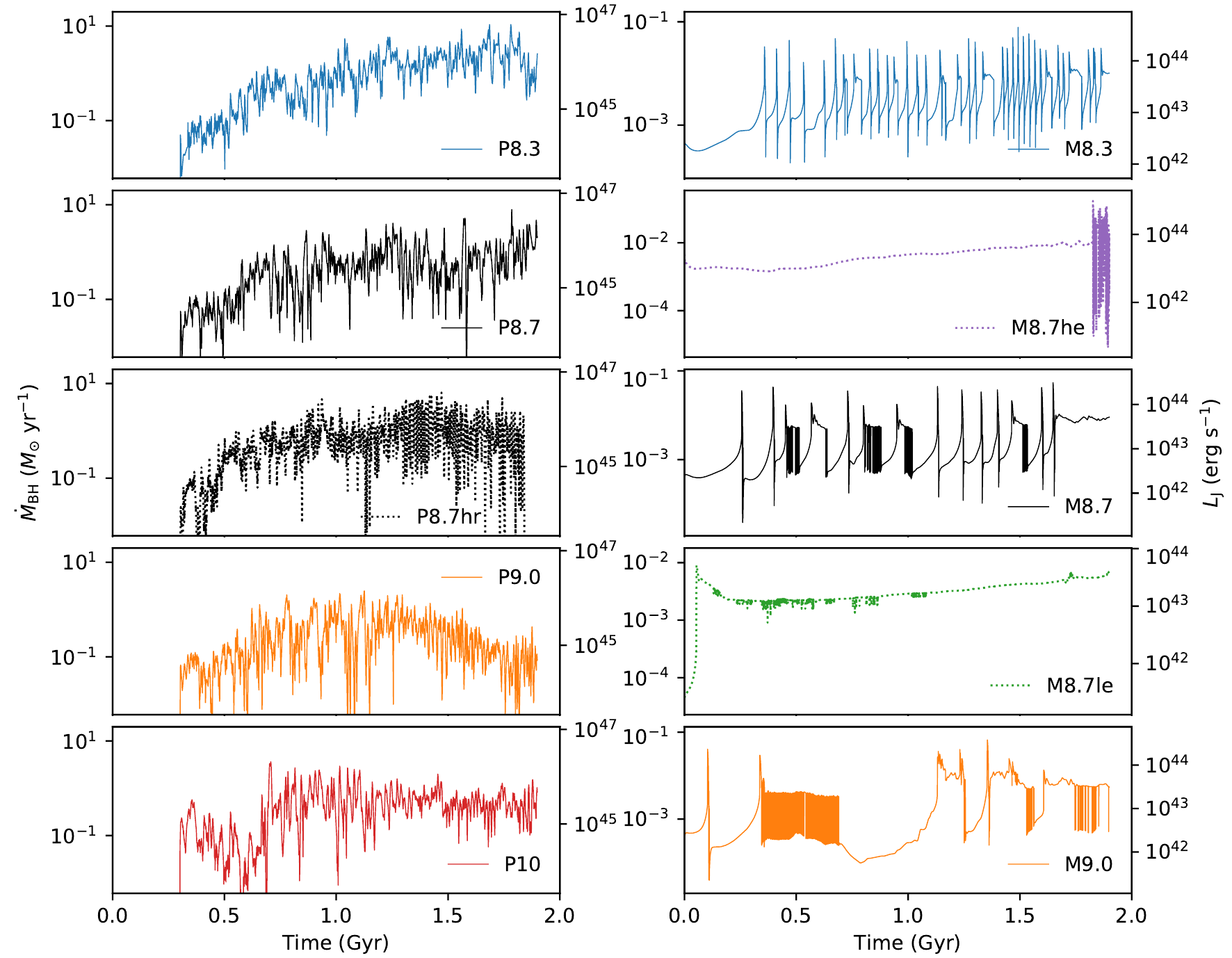}
\caption{Evolution of the SMBH accretion rate $\dot{M}_{\rm BH}$ and the corresponding AGN jet luminosity $L_{\rm J}$ over 2\,Gyr. The left panels summarizes the Perseus simulations, where $\dot{M}_{\rm BH}$ ($L_{\rm J}$) increases in the first 1\,Gyr before flattening or decreasing in the subsequent gigayear. Simulations of the M87 system is shown in the right panels, whose $\dot{M}_{\rm BH}$ ($L_{\rm J}$) in most cases show episodes of outburst on cycles of $\sim10-100\,{\rm Myr}$. In the case where $\varepsilon_{\rm acc}$ deviates from $10^{-2}$ (M8.7he and M8.7le), the cyclic behavior is broken, and the evolution is relatively level.}
\label{fig:AGN}
\end{figure*}

Properties of the hot plasma are characterized by the spherically symmetric profiles of electron number density $n_e$, temperature $T$, and metallicity $Z$, which are:
\begin{equation}
\begin{split}
n_e(r)&=0.22\left[1+\left(\frac{r}{0.93\,{\rm kpc}}\right)^2\right]^{-0.99/2}\,{\rm cm}^{-3},\\
k_{\rm B}T(r)&=3\,\frac{1+\left(\frac{r}{19\,{\rm kpc}}\right)^{1.9}}{3/1.55+\left(\frac{r}{19\,{\rm kpc}}\right)^{1.9}}\,{\rm keV},\\
Z(r)&=\left[0.3+0.7\left(\frac{r+50\,{\rm kpc}}{50\,{\rm kpc}}\right)^{-2.2}\right]\,Z_\sun.
\end{split}
\end{equation} 
These profiles are based on X-ray observations of M87 within 50\,kpc \citep[assuming $0.078\,{\rm kpc\,arcsec}^{-1}$;][]{Churazov2008, Russell2015}. At $r\gtrsim50\,{\rm kpc}$, out to the simulation boundary at 250\,kpc, the temperature profile flattens to a plateau of 3\,keV, in line with the measurements of the Virgo cluster \citep{Simionescu2017}. This also helps the plasma achieve hydrostatic equilibrium with the background gravitational potential described above. {We do not initialize the plasma with angular momentum or bulk motion.} The metallicity is set to $0.3\,Z_\sun$ at large radii, common to cluster outskirts \citep{Urban2017}, and consistent with the upper bound found in the Virgo cluster \citep[$0.22-0.32\,Z_\sun$;][]{Simionescu2017}.\footnote{Note in many works and also here $Z_\sun\approx0.02$, as found in \citet{Anders1989}, but the solar abundance has more recently been updated to 0.0134 in \citet{Asplund2009}.} {Based on these profiles, the gas is initialized as fully ionized plasma of hydrogen, helium, and electron species} (H\,\textsc{i}, H\,\textsc{ii}, He\,\textsc{i}, He\,\textsc{ii}, He\,\textsc{iii}, and $e^-$), {with an additional density field that represents metals. The plasma is allowed to cool radiatively throughout the simulation, down to $T\sim10\,{\rm K}$. The ionization state of the gas is updated in each timestep.} Further descriptions of the cooling processes modeled can be found in \citet{Qiu2018}.

\section{Results} \label{sec:results}

In an inside-out fashion, we use the following three subsections to present our simulation results on ($i$) the SMBH accretion rate and AGN luminosity evolution, ($ii$) the outflow properties at $r<20\,{\rm kpc}$ driven by jets, and ($iii$) the thermal properties of the X-ray emitting plasma regulated by AGN feedback. 

\subsection{AGN Evolution} \label{sec:AGN}

Given that we initialize our simulated galaxies and clusters with fully ionized hot plasma, the initial SMBH accretion rate is low and contributed entirely by the Bondi accretion of the hot gas. After $\sim10-100\,{\rm Myr}$, as the core region cools radiatively and cold gas forms {($T<3\times10^4\,{\rm K}$)}, $\dot{M}_{\rm BH}$ increases and is powered mainly by the assumed infall of cold gas. This triggers AGN outbursts that heat the core plasma, which in turn reduce the SMBH accretion. As a consequence, the SMBH accretion rate and AGN luminosity experience self-regulating cycles, as shown in Figure~\ref{fig:AGN}.

\begin{figure*}[t!]
\centering
\includegraphics[width=180mm]{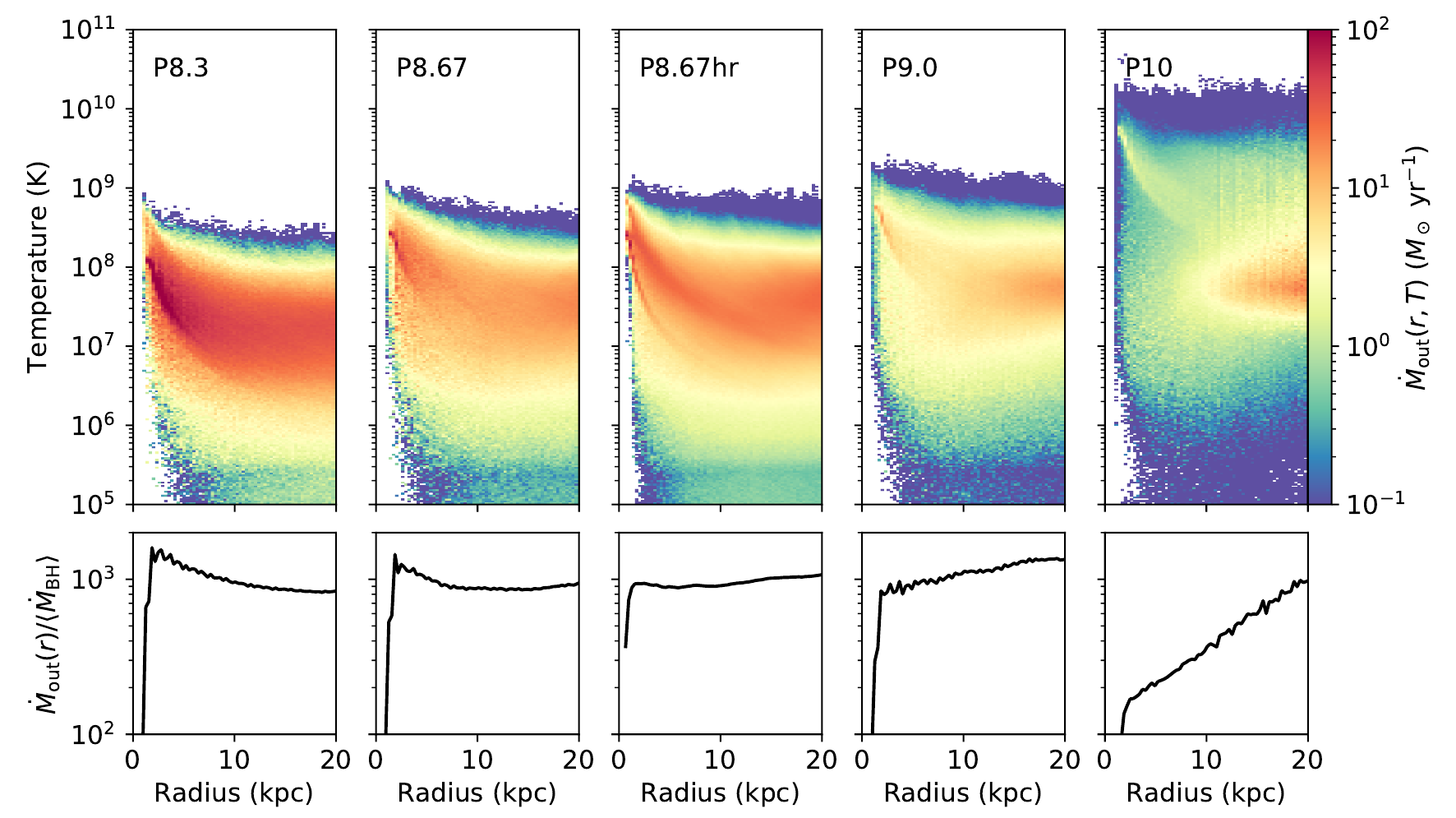}
\caption{{\it Top}: Radial temperature distribution of the plasma in Perseus simulations, color-coded with the measured mass outflow rate $\dot{M}_{\rm out}$ in each $(r,\,T)$ bin. Only gas with radial velocity $v_r>300\,{\rm km\,s}^{-1}$ is considered part of the outflow that counts toward the mass rate calculation. Each panel is the average of 100 snapshots between 1-2\,Gyr. Labels in the top-left corners indicate the simulation ID. {\it Bottom}: Total mass outflow rate (integrated over temperature from the corresponding top panels) as a function of radius, $\dot{M}_{\rm out}(r)$, normalized by the average black hole accretion rate, $\langle\dot{M}_{\rm BH}\rangle$, between $t=1-2\,{\rm Gyr}$ of each Perseus simulation (see Table~\ref{tab:params}).}
\label{fig:outflow}
\end{figure*}

In the case of Perseus, the first AGN outburst is triggered at $t\approx300\,{\rm Myr}$ due to the formation of dense cold gas around the central SMBH. While the SMBH accretion rate, or equivalently, AGN luminosity, shows short cycles on timescales of $10-100\,{\rm Myr}$, there is a common trend for all Perseus simulations: In the first 1\,Gyr, the plasma evolution is dominated by cooling, and $\dot{M}_{\rm BH}$ increases up to $\sim1\,M_\sun\,{\rm yr}^{-1}$ due to the accumulation of cold gas. After 1\,Gyr, $\dot{M}_{\rm BH}$ becomes relatively level for most simulation runs. P9.0, distinctively, shows a decrease of AGN luminosity after the peak around 1 Gyr. This is helped by the efficient heating from a hotter (faster) jet compared to P8.3 and P8.7(hr). However, the decrease in $L_{\rm J}$ does not transfer to the hottest jet modeled in P10. Due to the high outflow speed, a larger fraction of the feedback energy escapes the core region before being deposited to the plasma. Nevertheless, compared to P8.3 and P8.7, the average SMBH accretion rate is still lower in P10. For a numerical comparison, we provide the average accretion rate at
$t > 1\,{\rm Gyr}$ for the Perseus simulations in Table~\ref{tab:params}, ranging from $0.40-2.14\,M_\sun\,{\rm yr}^{-1}$. As a general trend, hotter (faster) jet leads to lower $\langle\dot{M}_{\rm BH}\rangle$ in Perseus simulations. We will later show in Figure~\ref{fig:cold} that this is due to an overproduction of cold gas in cooler (slower) jet simulations. 

$\dot{M}_{\rm BH}$ in the less massive system M87 however takes a different evolution path: After the initial cooling-dominated episode of a few hundred million years when $\dot{M}_{\rm BH}$ increases steadily, a short outburst of $L_{\rm J} >10^{44}\,{\rm erg\,s}^{-1}$ is able to reduce the accretion rate below $10^{-3}\,M_\sun\,{\rm yr}^{-1}$, which signals the start of a new AGN cycle. This is consistent with the measurements of M87 by \citet{Forman2017}, who find the outburst that took place $11-12\,{\rm Myr}$ ago injected $5-6\times 10^{57}\,{\rm erg}$ of energy within $1-3\,{\rm Myr}$, which also implies a peak jet luminosity $\sim10^{44}\,{\rm erg\,s}^{-1}$. 

The duration of the cycles depends on the particular feedback model employed, {because faster jets lead to a higher plasma temperature, which takes a longer time to cool through bremsstrahlung radiation ($\Lambda_{\rm B}\propto n^2\,T^{0.5}$)}. In M8.3 where jets are slower, the cycles are each around or below 50\,Myr. With the intermediate jet speed in M8.7, the cycles are around 100\,Myr. In M9.0, the cycle length is further extended, in some cases beyond 400\,Myr. {The duty cycle when the AGN jet is operating at high luminosities ($L_{\rm J}\gtrsim10^{43}\,{\rm erg\,s}^{-1}$) in these three cases is between $f_{\rm duty}=0.24-0.51$, as shown in Table~\ref{tab:params} and analyzed in Appendix~\ref{app:duty}. For completeness, we also infer the radiative output of the AGN at low accretion rates, and calculate the fraction of time AGN radiative luminosity $L_{\rm R}>10^{43}\,{\rm erg\,s}^{-1}$, which is between $f_{\rm R43}=3.2-6.6\%$.}

A change in the value of the accretion efficiency $\varepsilon_{\rm acc}$, however, breaks the episodic behavior. In both M8.7le and M8.7he, where $\varepsilon_{\rm acc} = 10^{-3}$ and $10^{-1}$, the accretion rate evolves smoothly throughout the simulation (with the exception of a short oscillation phase towards the end of the high efficiency run, triggered by an accumulation of cold gas, see also Figure~\ref{fig:cold}). In M8.7le, due to the low accretion efficiency $\varepsilon_{\rm acc}$, {more cold gas accumulates at the center before feeding the SMBH}. On the other hand, in M8.7he where the coupling is more efficient, $\dot{M}_{\rm BH}$ is dominated by the hot Bondi accretion. Interestingly, even with both jet modeling and accretion efficiency values, the average SMBH accretion rates over each 2\,Gyr simulation do not differ significantly from one another, with $\langle\dot{M}_{\rm BH}\rangle$ in the range $0.0033-0.0051\,M_\sun\,{\rm yr}^{-1}$, as summarized in Table~\ref{tab:params}. This is consistent with the picture that AGN feedback is in a dynamic equilibrium that compensates for the radiative cooling of the hot plasma.

\subsection{Outflow Properties} \label{sec:outflow}

In order to understand the thermal properties of the jet-driven outflows, as well as to provide a sanity check for the jet implementation, in this subsection we plot and examine the radial temperature distribution of the outflowing gas in our simulations. Outflows in M87, due to the short duration of outbursts, cannot necessarily be captured at their peak with evenly-spaced output snapshots. We therefore focus our examination on the second half of Perseus runs, given the relatively level evolution in jet luminosity. 

For each simulation snapshot, the inner $0.5-20\,{\rm kpc}$ is divided linearly into 64 radial shells, with the gas temperature divided on log scale into 256 bins between $10^5-10^{11}\,{\rm K}$. For each gas cell, $4\pi r^2\rho_i v_{r,i} V_i/V_r$ is summed over to indicate the outflow rate in a particular ($r,T$) bin, where $\rho_i$, $v_{r,i}$, and $V_i$ are respectively the density, radial velocity, and volume of a gas cell, and $V_r$ is the total volume for a given radial shell. In order to focus on global outflows driven by jets, rather than local turbulent motion, we only consider gas cells with outflow velocity $v_{r,i}>300\,{\rm km\,s}^{-1}$. The calculation is performed for 100 data outputs between 1-2\,Gyr, and then averaged to extract the outflow trend in each simulation. The resulting temperature distribution, color-coded with mass outflow rate, $\dot{M}_{\rm out}$, is shown in the top panels of Figure~\ref{fig:outflow}. {Note that unlike $\dot{M}_{\rm jet}$ which is injected by the AGN feedback model, $\dot{M}_{\rm out}$ is larger and contains more gas phases after the jet interacts with the ambient gas. In the bottom panels of the figure, we show the ratio between the total mass outflow rate and the average black hole accretion rate, $\dot{M}_{\rm out}/\langle\dot{M}_{\rm BH}\rangle$, as a function of radius. This ratio is around $10^2-10^3$ at the core, and is smaller when jets are launched with a lower mass-loading factor $m$. As the outflows evolve to larger radii, $\dot{M}_{\rm out}/\langle\dot{M}_{\rm BH}\rangle$ values all asymptote to $\sim10^3$ at 20\,kpc, closely balancing the inflow rate of the inward moving plasma, regardless of the feedback model employed. }

The maximum gas temperature is in general agreement with $T_{\rm out}$ in each simulation, which rises as the outflow speed increases. {Meanwhile, $\dot{M}_{\rm out}$ becomes more spread-out over temperature.} The color-coding in these plots reveal a main outflow stream that starts at the center around $T_{\rm out}$, {consistent with the AGN feedback model implemented. The main outflow stream} then cools both radiatively and due to mixing with the ambient medium as it rises to larger radii, up to $\sim 10\,{\rm kpc}$. Between $10-20\,{\rm kpc}$, the bulk of the outflow component stays between $10^7-10^8\,{\rm K}$, which becomes indistinguishable between different models. Observationally, it is also challenging to distinguish such outflows from the ambient intracluster medium in the absence of X-ray data with high angular and spectral resolution. However, the thermal properties within 10\,kpc, given the distinct features shown in these plots, can be utilized to constrain how AGN feedback operates. We therefore move on to the thermal properties of the hot plasma in the next subsection.

\subsection{Thermal Properties of the Hot Plasma} \label{sec:plasma}

\begin{figure*}[t!]
\centering
\includegraphics[width=180mm]{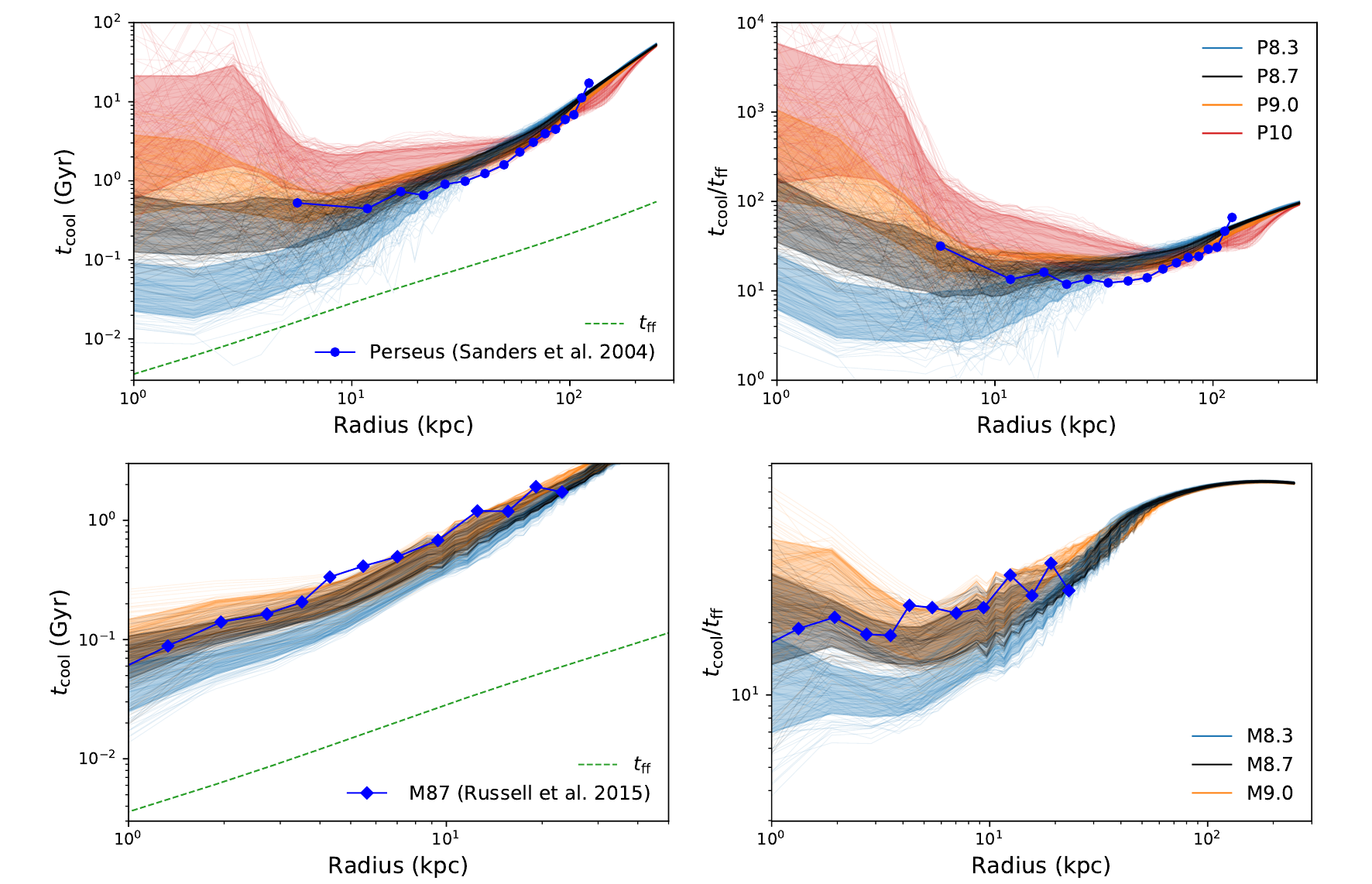}
\caption{A collection of cooling time and cooling-to-freefall timescale ratio profiles for the standard resolution (0.49\,kpc) runs in this work. Thin solid lines are taken from 100 snapshots between $1-2\,{\rm Gyr}$. The widths of the colored bands represent the $\pm \sigma$-span from the mean profile in each simulation in log scale. Profiles derived from observations are also over-plotted for comparison \citep{Sanders2004, Russell2015, McNamara2016}.}
\label{fig:profiles}
\end{figure*}

One feature common to cool-core clusters is that their radiative cooling time,
\begin{equation}
 t_{\rm cool}=\frac{n\,k_{\rm B}\,T}{(\gamma-1)\Lambda_X},
\end{equation}
is below 1\,Gyr within the central 10\,kpc \citep{Hudson2010, Hogan2017}, where $n$ is the plasma number density, and $\Lambda_X$ is the X-ray emissivity, {dominated by bremsstrahlung radiation $\Lambda_{\rm B}\propto n^2\,T^{0.5}$}. This is often compared with the freefall timescale, $t_{\rm ff}=\sqrt{2r/g}$, where $g$ is the gravitational acceleration. Theoretically, when the cooling-to-freefall timescale ratio drops below a certain threshold $\approx10$, the collapsing plasma will condense and form cold gas before falling into the center \citep{McCourt2012, Sharma2012}. Values of $t_{\rm cool}/t_{\rm ff}<10$ are however uncommon in systems with detections of cold molecular gas \citep{Pulido2018}, indicating that {\it in situ} cold gas formation may not be the dominant channel. 

In our recent work, we showed that cold gas forms from the warm ($T\lesssim10^7\,{\rm K}$) component of the AGN-driven outflows \citep{Qiu2020}, rather than from the thermal instability of the intracluster medium. {The minimum $t_{\rm cool}/t_{\rm ff}$ ratio most of the time stays around 20 in our simulation, in agreement with observations \citep[between $10-20$,][]{Hogan2017, Pulido2018, Olivares2019}}. One implication of this cold gas formation mechanism is that, {in the vicinity of the central AGN where multiphase gas coexists, even if some cold gas is heated to temperatures $T\lesssim10^7\,{\rm K}$ in the outflows, it may still be compressed and cool rapidly}, leading to star formation in an outflow \citep{Maiolino2017}, which constitutes a positive feedback channel. This also raises the question: which outflow component is heating the hot plasma?

We therefore examine the $t_{\rm cool}$ and $t_{\rm cool}/t_{\rm ff}$ profiles in our simulations with outflows dominated by different components at $T_{\rm out}>10^8\,{\rm K}$. {This quantity is chosen because: ({\it i}) It combines the thermal energy density ($\propto n T$) and the radiative cooling rate ($\propto n^2 T^{0.5}$; per unit volume) to estimate the cooling timescale ($\propto n^{-1} T^{0.5}$) from hot to cold gas phases, which may play a more fundamental role in how AGN regulates the thermal state of the ICM\footnote{{Another quantity with similar scalings is the specific entropy ($k_{\rm B} T n^{-2/3}$), which has similar radial profiles among different clusters and a common threshold below which cold gas tends to form \citep[e.g.,][]{Ponman1999, Tozzi2001, Ponman2003, Voit2005, Pratt2006, Donahue2006, Cavagnolo2008, Cavagnolo2009, Panagoulia2014, Babyk2018}. For brevity we do not compare the entropy profile in this work, and note that it gives similar information as $t_{\rm cool}$ \citep[e.g.,][]{Qiu2020}.}}. ({\it ii}) Compared with density and temperature profiles that may depend on the mass assembly of individual clusters, $t_{\rm cool}$($/t_{\rm ff}$) has a more universal floor \citep[$t_{\rm cool}\gtrsim 10^8\,{\rm yr}$, $t_{\rm cool}/t_{\rm ff}\gtrsim 10$, e.g.,][]{Hogan2017, Pulido2018, Olivares2019, Babyk2019} that can be used to constrain AGN feedback models. For completeness, in Appendix~\ref{app:profiles} we provide additional density and temperature profiles of the simulated clusters.}

In order to facilitate the comparison with existing X-ray data, the X-ray luminosity $L_X$ is taken to be the $0.1-10\,{\rm keV}$ X-ray thermal emissivity of the hot plasma. Finally, because some lower temperature plasma is adjacent to the cold gas that forms in the cluster, leading to non-radiative cooling through mixing \citep{Fabian2006}, as well as scattering and absorption, we remove the gas cells that are immediate neighbors with the cold gas \citep[similar to][]{Qiu2020}. This has minuscule effects on the M87 simulations due to a smaller amount of cold gas, but reduces the scatter in Perseus profiles with finely sliced radial shells. The similarities between Perseus and M87 simulations, as well as between low- and high-resolution runs (P8.7 and P8.7hr; see Appendix~\ref{app:resolution}) show that by removing the mixing layer, regardless of its thickness, the thermal profiles of the hot plasma are {primarily determined by the jet mass-loading factor $m$}.

Figure~\ref{fig:profiles} shows the $t_{\rm cool}$ and $t_{\rm cool}/t_{\rm ff}$ profiles for Perseus and M87 runs with standard resolution (0.49\,kpc) in 100 snapshots between $1-2\,{\rm Gyr}$. In both systems, profiles of the central region differ significantly based on the outflow model employed: For $T_{\rm out}=10^{8.33}\,{\rm K}$ in P(M)8.3, the central cooling time is almost always below $10^{8}\,{\rm yr}$. The minimum $t_{\rm cool}/t_{\rm ff}$ also spends a significant amount of time below 10, in apparent contrast with observed profiles of giant galaxies in clusters \citep{Hogan2017}. This indicates that the particular feedback model in P(M)8.3 has exaggerated the mass-loading factor of jets, resulting in a slow outflow incapable of heating the intracluster medium. On the other extreme, an overly fast outflow in simulation P10 drives the central cooling time above 1\,Gyr, which is no longer considered a cool-core cluster in this context. While this provides a mechanism for cool-core clusters to transition to a non-cool-core, overly light outflows with $v_{\rm out}>10,000\,{\rm km\,s}^{-1}$ is also unlikely the main heating channel in these systems.

The outflows with intermediate speeds and temperature, P(M)8.7 and P(M)9.0, are in good agreement with the observed profiles of Perseus and M87. In particular, in both systems, the central profiles is better described by the P(M)8.7 simulations, suggesting that the mass loading factor is on the order of hundreds. Beyond a few kiloparsecs, however, P(M)9.0 is in better agreement with the observed profiles. This, on the one hand, may indicate a smaller mass-loading factor in previous AGN outburst in Perseus and M87. A more likely scenario, however, is that the dissipation of outflow energy in real clusters differs from what is modeled in our idealized hydrodynamic simulations. 

Intriguingly, a higher outflow temperature does not necessarily result in a longer $t_{\rm cool}$ at large radii. While the profiles are similar outside of 50\,kpc, which is expected given the long cooling time of the plasma in the outskirts, the trends at large radii are reversed among outflow models with different $T_{\rm out}$, in both Perseus and M87 simulations. In the outskirts, hotter outflow models lead to lower $t_{\rm cool}$ profiles. This suggests inefficient outskirt heating by a lighter, hotter jet, which can be attributed to two factors: ($i$) By employing a lighter jet, the central AGN may be less efficient at driving sound waves to transport energy to cluster outskirts. ($ii$) As shown by \citet{Qiu2020}, hotter outflows experience more ram pressure deceleration from the intracluster medium, decreasing their spatial reach. An unintuitive implication of this phenomenon is that slower (cooler) outflows play a more important role in heating the cluster outskirts. 

In addition to the standard simulations presented in Figure~\ref{fig:profiles}, we have also conducted the same analysis for simulations with higher resolution (P8.7hr) or different accretion efficiencies ($\epsilon_{\rm acc}$; M8.7he, M8.7le). In light of their modest differences from the standard run results, we show profiles of these simulations in the Appendix, and proceed to discuss the implications and conclusions of our work in the next section.

\section{Discussion} \label{sec:discussion}

\subsection{Implications for AGN Physics}\label{sec:implication}
In this work, by varying the mass-loading factor of AGN-driven outflows in our simulations, we obtained thermal profiles of the hot plasma that are systematically distinct from each other for the inner $\sim10\,{\rm kpc}$. This allowed us to compare the resulting profiles with X-ray observations of the intracluster medium, and constrain the mass-loading factor between $200-500$ around 1\,kpc, the nominal jet-launching radius in our simulations. Equivalently, this implies that the bulk flow is moving with speeds between $4,000-7,000\,{\rm km\,s}^{-1}$, or that the thermalized outflow temperature is between $10^{8.67}-10^{9}\,{\rm K}$. 

{Considering a dynamical balance of the inflows and outflows near the SMBH, the mass-loading factor $m>100$ may be closely related to the inverse of the accretion efficiency $\varepsilon_{\rm acc}^{-1}$. Motivated by observations \citep{Kuo2014, Russell2015}, $\varepsilon_{\rm acc}$ is taken to be $10^{-2}$ in most of the simulations (see Appendix~\ref{app:resolution} for a variance of this parameter.). This means the inflow accumulates at a rate $100\times \dot{M}_{\rm BH}$. The AGN-driven outflows then need to ({\it i}) counter the mass accumulation dictated by the gravity of the SMBH and the stellar bulge, and ({\it ii}) compensate for the radiative loss of the gas energy to maintain the thermal structure. The ratio $m/\varepsilon_{\rm acc}^{-1}=2-5$ therefore represents the outflow rate needed per unit mass inflow rate to strike a balance, especially at the peaks of AGN activity, when most of the feedback energy is released. This results in a low effective accretion rate for SMBHs residing in giant elliptical galaxies, meanwhile restoring the thermal support of the hot plasma consistent with X-ray observations. On the other hand, in simulation P10 where $m\approx20$, much less than the 100 needed to achieve the dynamical balance between inflow and outflows, most of the feedback energy is stored in thermal form. This transforms the core thermal structure to a state that inhibits the direct infall of the cooling flow, but is inconsistent with observations of the hot plasma in giant elliptical galaxies.}

{Another important question to consider is the origin of the baryons in the outflows. Simulations in this work do not resolve the plasma evolution below 1\,kpc, where the bulk of the baryonic outflows originate. Theoretically, the fast baryons can be driven by relativistic AGN jets, which implies a significant transport of momentum from the pair plasma to the surrounding baryonic medium. On the other hand, it is also possible that the baryons are directly launched as disk winds from the accretion flow. Therefore, we discuss these two mass-loading scenarios in the following paragraphs.}

(1) If the outflows are driven primarily by relativistic jets, this implies a significant baryonic mass entrainment from a few gravitational radii to $\sim1\,{\rm kpc}$. This latter radius is consistent with models of stellar mass entrainment, which find that initially relativistic jets will be decelerated to non-relativistic speeds within $1-2\,{\rm kpc}$ \citep{Hubbard2006, Perucho2014} when jet luminosity $L_{\rm J}\lesssim10^{43}\,{\rm erg\,s}^{-1}$ (or equivalently, $\dot{M}_{\rm BH}\lesssim2\times10^{-3}\,M_\sun\,{\rm yr}^{-1}$ assuming $\eta=0.1$). In this scenario, interstellar medium such as stellar wind is the main supplier of baryons. {In the case of nearby elliptical galaxies, the stellar mass loss rate for the central kiloparsec is $\sim0.1\,M_\sun\,{\rm yr}^{-1}$ \citep{Padovani1993, Ho2009}. With a mass-loading factor $m\sim100$, this also implies that stellar mass loss can sustain the baryonic mass-loading up to $\dot{M}_{\rm BH}\sim10^{-3}\,M_\sun\,{\rm yr}^{-1}$. Beyond this rate, the SMBH accretion is likely fueled by the cold gas that drops out of the hot plasma (e.g., the concurrent peaks of $\dot{M}_{\rm BH}$ and $M_{\rm cold}$ in M87 simulations, shown in Figures~\ref{fig:AGN} and \ref{fig:cold}). At these regimes, cold gas clumps may greatly supplement the baryons to the outflows.} Judging from the agreement between our models and X-ray observations of the plasma properties in Perseus A and M87, $m$ on the order of hundreds seems to be a universal factor for giant elliptical galaxies, regardless of their halo mass.

{
(2) Another possible scenario is that the outflows are driven primarily by disk winds from the accretion flow \citep[e.g.,][]{Yuan2018, Yoon2018, Cui2020, Yang2021, Shi2021}. 
In radiatively inefficient accretion flows, the gas inflow rate scales with radius as
\citep{Blandford1999, Yuan2014}:
\begin{equation}
\begin{split}
 \dot{M}_{\rm in}(r) &= \dot{M}_{\rm BH} \left(\frac{r}{R_{\rm in}}\right)^{s}\ {\rm for}~r \geq R_{\rm in}\\
\end{split}
\end{equation}
where the index of $s$ is close to unity for MHD accretion flows \citep[see e.g.,][]{Yuan2015},
the innermost radius is set to $R_{\rm in}= f\,r_{\rm S}$, $r_{\rm S}=2\,G\,M_{\rm BH}/c^2$ is the Schwarzschild radius, 
$G$ is the gravitational constant, $f\sim O(10-100)$ is a dimension-less factor obtained from numerical simulations.
Above this radius, there is an outflow that scales similarly with radius; namely $\dot{M}_{\rm in}(r) \approx \dot{M}_{\rm jet}(r)$ 
\citep[e.g.,][]{Stone1999, Quataert2000, Inayoshi2018}.
{Supposing that a large fraction of the wind mass is launched within a characteristic radius at an outflow velocity of $v_{\rm out}$, the scale is estimated as $R_{\rm w} \simeq 2 GM_{\rm BH}/v_{\rm out}^2= r_{\rm S}(v_{\rm out}/c)^{-2}$.}
Therefore, the mass-loading factor is given by 
\begin{equation}
\begin{split}
m &= \frac{\dot{M}_{\rm jet}(R_{\rm w})}{\dot{M}_{\rm BH}} \approx \left(\frac{R_{\rm w}}{R_{\rm in}}\right)^{s}\\
&\approx 360~\left(\frac{f}{10}\right)^{-1}\left(\frac{v_{\rm out}}{5,000~{\rm km~s}^{-1}}\right)^{-2},
\end{split}
\end{equation}
where $s=1$ is assumed. The value is consistent with our numerical result of $m=200-500$. {We caution, however, this argument depends on the inner radius where the disk wind ceases and the mass inflow rate becomes constant toward smaller radii, i.e., the location of $R_{\rm in}$ (or equivalently $f$), which may change the detailed energy transport near the BH event horizon.} With current knowledge, both wind and jet mechanisms seem able to power the outflows required to reconstruct the thermal profiles of the hot plasma.
}

Given that the mass-loading of AGN-driven outflows depends heavily on the material or accretion flow below 1\,kpc, we emphasize that the constraint found for the mass-loading factor, $m>100$, applies mainly to giant elliptical galaxies such as Perseus A and M87, {whose central 1-kpc stellar mass exceeds $2\times10^{10}\,M_\sun$ \citep{Mathews2006, Romanowsky2001}}. On the other hand, the elevated central plasma profiles in our simulations generated by a lighter outflow (e.g., P10 in Figure~\ref{fig:profiles}) may have an implication for AGN feedback in smaller systems. In less massive galaxies with smaller core stellar density, it follows that their jet mass loading will be limited, which further leads to a faster, hotter AGN outflow that drives the formation of hot plasma bubbles.  An example of such hot plasma is the $\gamma$- and X-ray bubbles in our Milky Way \citep{Su2010, Predehl2020}, {whose core stellar mass within 1\,kpc is $\sim6\times10^9\,M_\sun$ \citep[e.g.,][circular velocity $\sim160\,{\rm km\,s}^{-1}$]{McMillan2011}. This produces an ``umbrella'' of hot plasma that prevents the direct collapse of the cooling gas in less massive galaxies.}

\subsection{Cold Gas Properties}\label{sec:cold}


\begin{figure}[t!]
\centering
\includegraphics[width=86mm]{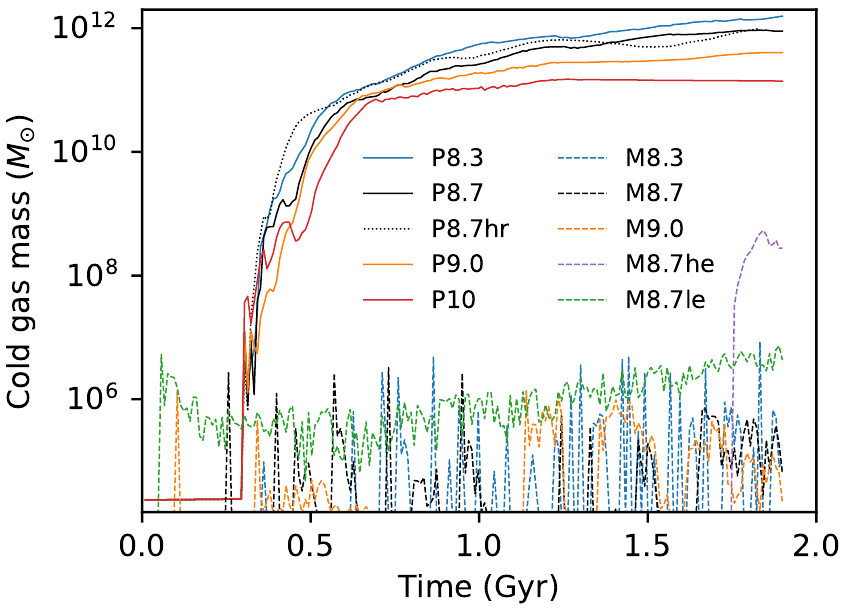}
\caption{Evolution of cold gas mass in each simulation run over 2\,Gyr, represented by the total amount of neutral hydrogen.}
\label{fig:cold}
\end{figure}

\begin{figure*}[t!]
\centering
\includegraphics[width=180mm]{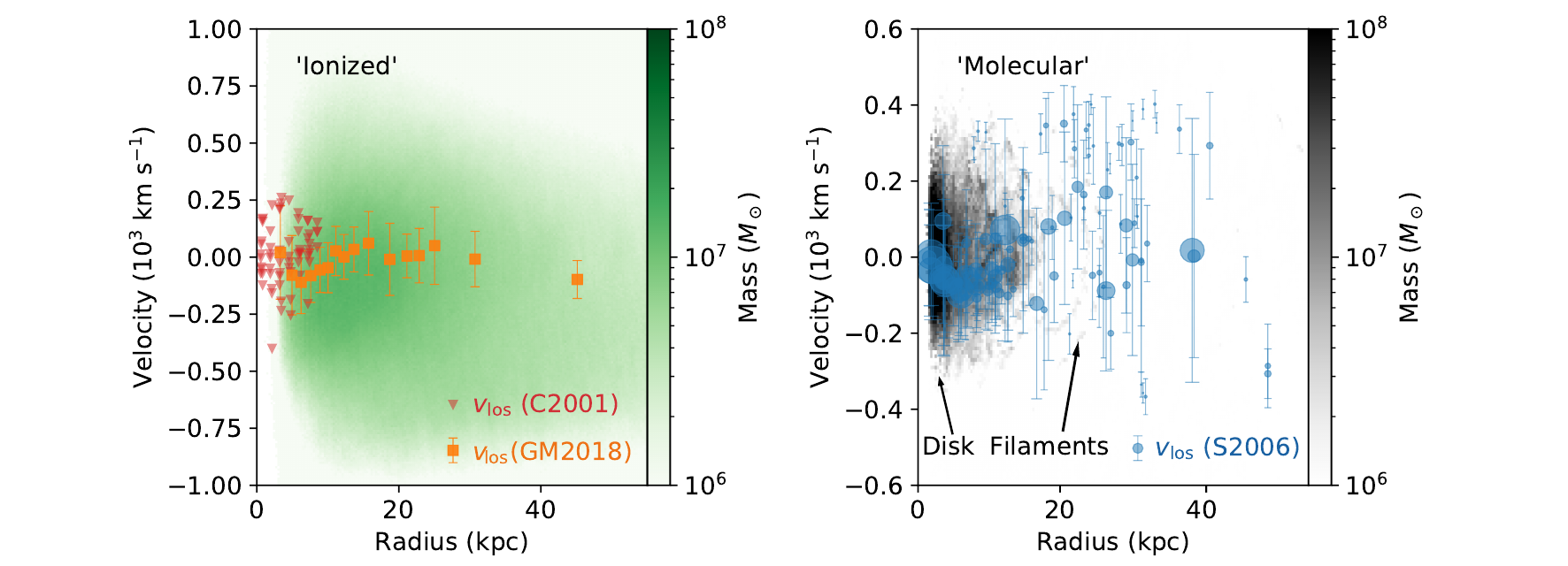}
\caption{{Radial velocity distribution of `ionized' ($10^4<T<10^5\,{\rm K}$) and `molecular' ($T<100\,{\rm K}$) components of the simulated Perseus cluster in P8.7hr, color-coded with mass (same as the middle panels in Figure~\ref{fig:vr}, with an extended radial range out to 55\,kpc). For comparison, {\it line-of-sight} velocity ($v_{\rm los}$) observations, as a function of {\it projected} radius in the Perseus system is overplotted \citep[][the systemic velocity is taken to be $5,264\,{\rm km\,s}^{-1}$ and subtracted from all three data sets]{Conselice2001, Salome2006, GM2018}. {\it Left}: The error bar represents the velocity dispersion of the H$\alpha$-emitting gas \citep{GM2018}.  {\it Right}: The marker size is proportional to the measured molecular gas mass \citep[ranging from $\sim10^7-10^9\,M_\sun$;][]{Salome2006}. The vertical spread of the error bar indicates CO(2-1) line width. Arrows point to the locations of the rotationally supported disk and the filaments.}}
\label{fig:vlos}
\end{figure*}

One criterion that has been widely used to assess the effectiveness of AGN jet modeling is the reduction in the amount of cold gas. Ideally, energy released by jets compensates for the radiative losses of the hot plasma, reduces the amount of cold gas, and quenches the formation of stars. In Figure~\ref{fig:cold} we examine the total amount of cold gas in our simulated Perseus and M87 in order to assess the heating efficiency in different models from the perspective of the cold gas production. The amount of neutral hydrogen is collected from each cell within $r<250\,{\rm kpc}$ to represent the cold gas mass. 

In the case of M87, the cold gas mass evolution in runs with standard accretion efficiency $\varepsilon_{\rm acc}=10^{-2}$ resembles that of the accretion rate. $M_{\rm cold}$ stays below $\sim10^4\,M_\sun$ most of the time, but this value is periodically breached to a peak of a few$\times10^6\,M_\sun$, which triggers an AGN outburst that in turn reduces $M_{\rm cold}$ back to low values. The peak values are within the upper limit found in searches of molecular gas \citep{Braine1993, Tan2008}. In the low-efficiency run M8.7le where $\varepsilon_{\rm acc}=10^{-3}$, $M_{\rm cold}$ stays relatively level around $10^6\,M_\sun$. M8.7he, on the other hand, did not gain a significant amount of cold gas until an outburst towards the end of the simulation. Overall, the amount of cold gas created in the simulations is in agreement with observations of M87.

The cold gas evolution in the Perseus simulations is however less contained. Being a more massive halo, radiative cooling of the hot plasma means a cooling rate of a few$\times100\,M_\sun\,{\rm yr}^{-1}$. By the end of the 2-Gyr simulation, {the cold gas settles into a rotationally supported disk (see also Figure~\ref{fig:vlos})}, and $M_{\rm cold}$ reaches $\gtrsim10^{12}\,M_\sun$, $1-2$ orders of magnitude larger than typical cold gas mass in cool-core clusters \citep{Pulido2018}. Even with the hottest outflow explored in this work \citep[P10, as implemented in many other works, e.g.,][]{Gaspari2012a, Li2014}, the reduction is only by a factor of 10. This could be attributed to two main causes: ($i$) As mentioned earlier, there is a mixing layer at the interface between the hot plasma and the cold gas that is cooling non-radiatively. In coarse resolution simulations where this is only resolved by one cell width, the cooling may be overestimated {(see Appendix~\ref{app:vr} for a discussion)}. However, a proper treatment of this process requires resolutions beyond the scope of this work. ($ii$) Additional physical processes such as thermal conduction may contribute to the heating of the intracluster medium \citep[][]{Voigt2004, Bogdanovic2009, Yang2015}. Nevertheless, the solution to the overproduction of cold gas in more massive halos, as we demonstrate in this work, does not lie with employing outflows with $v_{\rm out}>10^4\,{\rm km\,s}^{-1}$, due to their propensity to raise the plasma thermal properties beyond observational constraints. This is in agreement with our previous work which features a more ``gentle'' feedback implementation, with outflow speeds on the order of a few$\times1,000\,{\rm km\,s}^{-1}$ \citep{Qiu2018, Qiu2020}.

{
Another important constraint from observations is the line-of-sight velocity of the filamentary gas. For a direct comparison with the cold gas velocities in the Perseus system, in Figure~\ref{fig:vlos} we plot the {\it line-of-sight} velocity measurements as a function of {\it projected} radius from H$\alpha$ and CO observations \citep{Conselice2001, Salome2006, GM2018}, on top of the radial velocity distribution of the simulated gas components corresponding to the H$\alpha$ and CO emitters (`ionized': $10^4\,{\rm K}<T<10^5\,{\rm K}$ and `molecular': $T<100\,{\rm K}$; see also Appendix~\ref{app:vr}). Note that $v_{\rm los}<0$ does not necessarily mean the gas is flowing towards the central AGN. Compared with H$\alpha$ filaments probing $T\sim10^4\,{\rm K}$ gas, the `ionized' gas has a larger vertical spread, indicating that less massive components with higher velocities exist in the simulations. In reality, they are either too faint to be detected, or non-existent due to physical processes not captured in our modeling. When the gas cools below $100\,{\rm K}$, it is mainly dominated by molecular gas, which can be traced by CO lines, as shown in the right panel of Figure~\ref{fig:vlos}. Both the simulated `molecular' gas and the observations indicate that this component has relatively low velocities, and most of the gas is concentrated in the central 10\,kpc. However, in simulations the cold gas persists and settles into a rotationally supported disk, because the angular momentum does not cancel out completely \citep[e.g.,][]{Qiu2018}. {This is however uncommon in real clusters, indicating that even though cold filaments are spatially spread-out and common in cool-core clusters, each individual filament is short-lived, either becoming fuel for star formation \citep{Tremblay2015}, or reheated by subsequent stellar feedback and local processes, before settling into a disk.}
}

\section{Conclusions}\label{sec:conclusion}
In this work we use simulations of the AGN-driven outflows to explore their impact on the thermal properties of the hot plasma in the intracluster space. By varying the baryonic mass-loading factor of the jets, which also dictates the speed and thermalized temperature of the outflows, we compare the resulting thermal profile with X-ray observations of the intracluster medium. The main findings of our work is summarized below:

1. Regardless of the physical mechanism behind the AGN-driven outflows, be they relativistic jets or disk winds, the outflow-to-accretion mass-loading factor (calculated assuming the AGN feedback efficiency $\eta=0.1$) lies between $200-500$ by the time they extend to 1\,kpc from the SMBH. Outflows launched with this factor results in thermal profiles consistent with X-ray observations of the intracluster plasma in both Perseus and M87, the systems we model. Equivalently, this indicates that the outflow is launched at a speed between $4,000-7,000\,{\rm km\,s}^{-1}$, and the thermalized outflow temperature is between $10^{8.67}-10^{9}\,{\rm K}$. It also means that observationally the outflow rates are systematically underestimated because this component is undetectable without kinetically-resolved X-ray data of the central regions in elliptical galaxies. 

2. While the maximum observed baryonic outflow speed exceeds $10,000\,{\rm km\,s}^{-1}$ in many systems, it is unlikely that it represents the bulk of the flow in giant elliptical galaxies. Outflows modeled with such high speeds lead to an elevated cooling time profile within 10\,kpc, inconsistent with observed properties of Perseus A and M87. However, in less massive galaxies where jets couple inefficiently with the baryons, the outflows may retain high speeds, creating energetic bubbles that prevent the direct collapse of the cooling circumgalactic medium and the subsequent galaxy growth, such as the $\gamma$- and X-ray bubbles in our Milky Way. 

3. With the standard parameters shown above, AGN-driven outflows are capable of regulating the amount of cold gas in less massive halos such as M87. It however fails to reduce the formation of cold gas in massive halos like Perseus. Nevertheless, the consistency of the plasma properties between modeled and observed Perseus clusters indicates that the overproduction is not due to jet modeling, {but rather due to the lack of resolution or physical processes necessary to resolve the dense clumps, such as the mixing layer between hot and cold gas, as well as thermal conduction over steep temperature gradients inside the plasma.}

\vspace{5mm}

We thank the anonymous referee for useful comments and suggestions that helped improve this work. Y.Q. thanks Feng Yuan for useful discussions. This work is supported by the National Key R\&D Program of China (2016YFA0400702), the National Science Foundation of China (11721303, 11991052, 11950410493, 12003003, 12073003), the China Postdoctoral Science Foundation (2020T130019), and the High-Performance Computing Platform of Peking University. T.B. acknowledges support provided by the National Aeronautics and Space Administration through Chandra Award Number TM7-18008X issued by the Chandra X-ray Center, which is operated by the Smithsonian Astrophysical Observatory for and on behalf of the National Aeronautics Space Administration under contract NAS803060.


\software{Enzo \citep{Bryan2014}, 
          {yt \citep{Turk2011}}.
          }

\appendix

\section{Resolution and Accretion Efficiency Study}\label{app:resolution}

For the modeling of AGN-driven outflows, the primary parameter explored in this work is the mass-loading factor $m$. We therefore compare plasma properties resulting from different $m$ values in the main text above. Another parameter, the accretion efficiency $\varepsilon_{\rm acc}$, was already examined in \citet{Qiu2018} for the Perseus setup. Results from the early work showed that $\varepsilon_{\rm acc}\gtrsim10^{-2}$ leads to similar cluster evolution. Observationally, the accretion rate in M87 reduces from $\sim0.1\,M_\sun\,{\rm yr}^{-1}$ at the Bondi radius to $\lesssim10^{-3}\,M_\sun\,{\rm yr}^{-1}$ at 21 Schwarzschild radii, also indicating that $\varepsilon_{\rm acc}\approx 10^{-2}$. Out of an abundance of caution, we study the effect of this parameter again, for the M87 setup, with three values: $\varepsilon_{\rm acc}=10^{-3}$ for M8.7le, $10^{-2}$ for M8.7, and $10^{-1}$ for M8.7he. 

\begin{figure*}[h!]
\centering
\includegraphics[width=180mm]{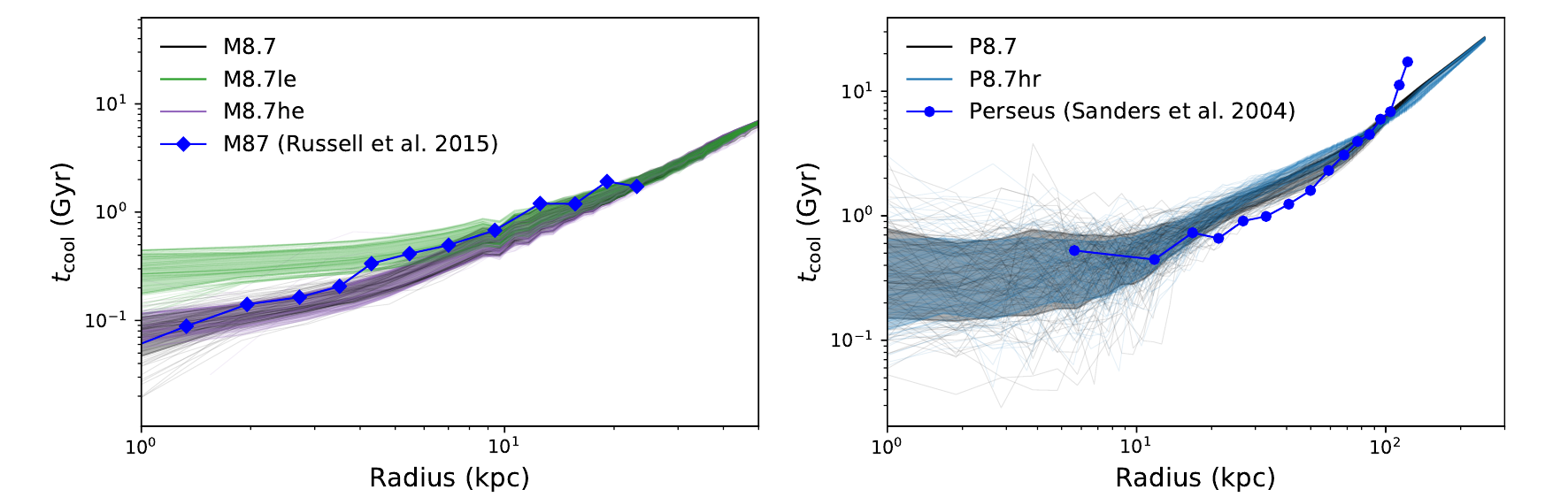}
\caption{A collection of cooling time profiles for the simulation runs with non-standard parameters in this work. Thin solid lines are taken from 100 snapshots between $1-2\,{\rm Gyr}$. The widths of the colored bands represent the $\pm \sigma$-span from the mean profile in each simulation in log scale. Profiles derived from observations are also over-plotted for comparison \citep{Sanders2004, Russell2015}.}
\label{fig:profiles2}
\end{figure*}

Cooling time profiles from these simulation runs are shown in Figure~\ref{fig:profiles2}. The profiles in M8.7 and M8.7he are similar and in agreement with observations of M87. M8.7le, on the other hand, has a higher central profile. {The low accretion efficiency in this run implies that inflows are accumulating at a rate $10^3\times\dot{M}_{\rm BH}$, larger than the rate at which they are expelled ($m\times\dot{M}_{\rm BH}$), which leads to a reservoir of cold gas at the center (Figure~\ref{fig:cold}). Meanwhile, the constant injection of outflows in M8.7le also increases (reduces) the temperature (density) of the plasma, ultimately reaching a steady state where the intracluster medium is cooling at a lower rate compared with simulation M8.7, while maintaining a central reservoir of cold gas that slowly feeds the SMBH. On the other hand, $\dot{M}_{\rm BH}$ in M8.7he also reaches a steady state, but the high accretion efficiency ensures there is no accumulation of cold gas, and the cooling time profile is not elevated compared to M8.7.}

In order to confirm the convergence with simulation resolution, we also performed a high-resolution run for the Perseus setup (0.24\,kpc; P8.7hr). Its accretion rate evolution, outflow properties, and total cold gas mass are all similar to P8.7, as shown in Figures~\ref{fig:AGN}, \ref{fig:outflow}, \ref{fig:cold}. The right panel in Figure~\ref{fig:profiles2} also reveals that the cooling time profiles in P8.7hr share similar average and standard deviation values with the low-resolution simulation P8.7 in the central region. The agreement in all of these properties indicates that the impact of the mass-loading factor found in the main text does not depend on simulation resolution. 

\section{Mass and Velocity Distribution of the Multiphase Gas}\label{app:vr}
{
Besides the thermal properties of the hot plasma, the mass and line-of-sight velocity measurements of the various gas components in giant elliptical galaxies also provide strong constraints on models of AGN feedback. In order to derive quantities that can be directly compared with observations, as well as highlight areas that will require further investigation, in this section we examine the radial mass and velocity distribution of the simulated Perseus cluster. For brevity we show in Figure~\ref{fig:vr} only properties derived from the high-resolution run P8.7hr, and note that plots from other simulations are qualitatively similar.

\begin{figure*}[h!]
\centering
\includegraphics[width=180mm]{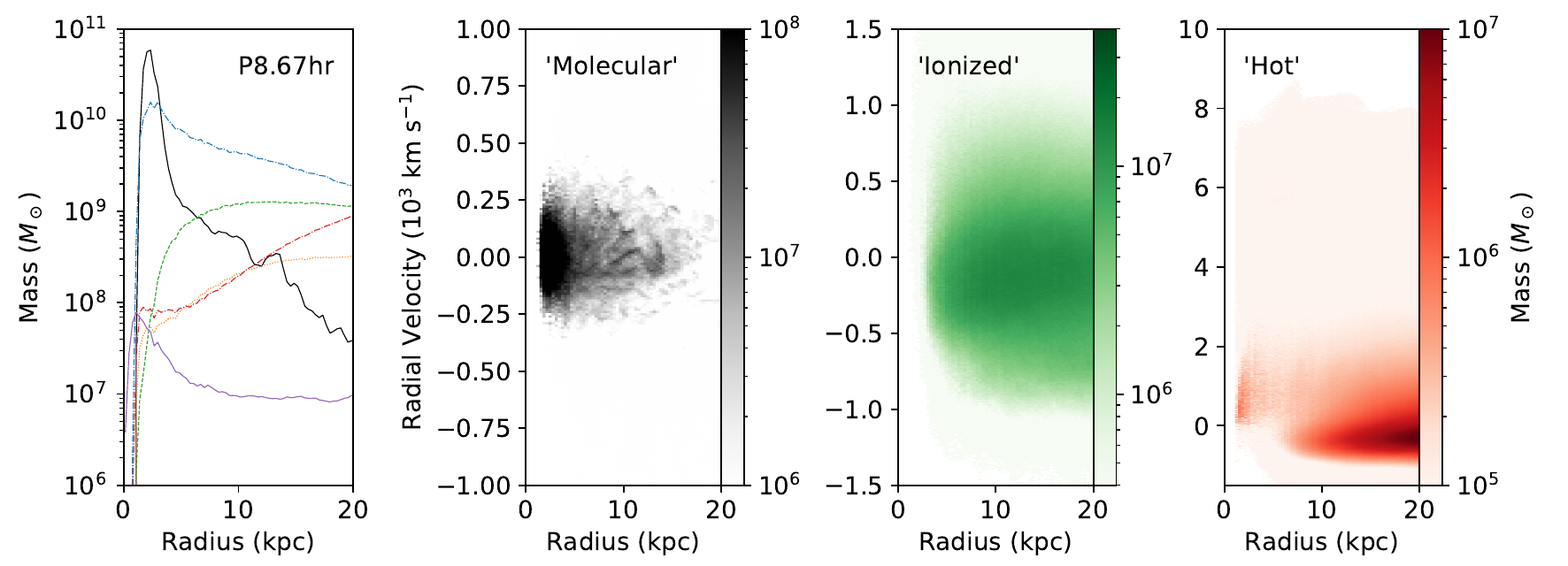}
\caption{Radial mass and radial velocity distribution of the gas in simulation P8.7hr. {The leftmost panel shows the enclosed mass of each radial shell of width $\sim0.3$\,kpc. Different colors represent gas within different temperature ranges:  $T<10^2\,{\rm K}$ (`molecular', black), $10^2\,{\rm K}<T<10^4\,{\rm K}$ (blue), $10^4\,{\rm K}<T<10^5\,{\rm K}$ (`ionized', green), $10^5\,{\rm K}<T<10^7\,{\rm K}$ (orange), $10^7\,{\rm K}<T<10^8\,{\rm K}$ (`hot', red), and $T>10^8\,{\rm K}$ (purple).} Intensity of colors in velocity distributions indicates the cumulative mass in each $(r,v_r)$ bin. Each panel is averaged over 100 snapshots between $1-2\,{\rm Gyr}$.}
\label{fig:vr}
\end{figure*}

The radial range between $0-20\,{\rm kpc}$ is divided into 64 spherical shells, yielding a shell width of $\approx0.3\,{\rm kpc}$. In each shell, we categorize the gas into six temperature ranges: $T<10^2\,{\rm K}$ (`molecular', black), $10^2\,{\rm K}<T<10^4\,{\rm K}$ (`cold', blue), $10^4\,{\rm K}<T<10^5\,{\rm K}$ (`ionized', green), $10^5\,{\rm K}<T<10^7\,{\rm K}$ (`warm', orange), $10^7\,{\rm K}<T<10^8\,{\rm K}$ (`hot', red), and $T>10^8\,{\rm K}$ (`extra-hot', purple). We then calculate the total mass for each temperature range in a spherical shell, and obtain the radial mass distribution for a given output snapshot. This distribution is then averaged over 100 snapshots between $1-2\,{\rm Gyr}$, and plotted in the first panel of Figure~\ref{fig:vr}. In this plot, the `hot' gas, which ranges between $10^8-10^9\,M_\sun$ per shell, is similar to the initial density distribution, due to the regulation of AGN feedback. The `extra-hot' component that represents the initial outflow injection is comparable to the `hot' gas near the launch site. Beyond a few kpc, this component drops greatly due to mixing and radiative cooling. The massive `molecular' component dominates in the central few kpc, in the form of a rotationally supported {disk \citep[e.g.,][see also Figure~\ref{fig:vlos}]{Qiu2018}}. The mixing layer, i.e., the `cold'+`ionized'+`warm' components, which is not resolved in our simulations, is also a significant mass component throughout this radial range. Considering that the cold gas is overproduced in simulations by a factor of $10-100$ relative to a real intracluster environment, in reality all of these components ($T\lesssim10^7\,{\rm K}$) likely stay at only $1-10\%$ of the shown values, surrounding the cold filaments, {such as the warm molecular hydrogen \citep[$T\sim2,000\,{\rm K}$,][]{Edge2002, Wilman2002, Hatch2005}, the O\,$\textsc{vi}$ line-emitting gas \citep[$T\sim10^{5.5}\,{\rm K}$,][]{Bregman2006}, and the soft X-ray gas \citep[$T\sim10^6\,{\rm K}$, e.g.,][]{Fabian2006}}. Resolving the mixing layer with much higher resolution, as well as including physical processes such as conduction that can be efficient in clusters with steep temperature gradients, is a key aspect for future investigation. 

\begin{figure*}[t!]
\centering
\includegraphics[width=180mm]{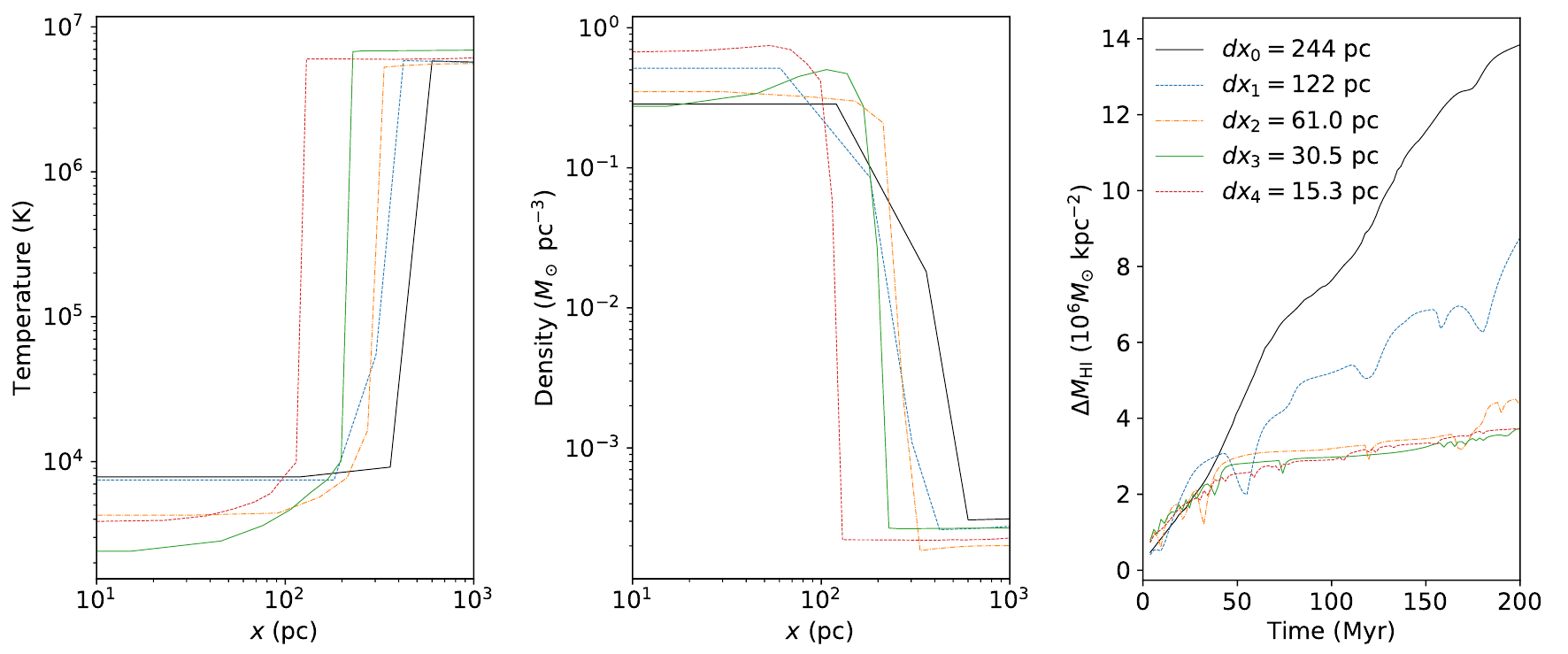}
\caption{Mass growth onto a $T\sim10^4\,{\rm K}$ cold gas sheet embedded in a 1\,keV plasma. First and second panels show the temperature and density profiles at $t=200\,{\rm Myr}$. The third panel shows the growth of neutral hydrogen mass per kpc$^2$. Different lines represent simulations with various resolution, as indicated by the legend.}
\label{fig:mix}
\end{figure*}

We note, however, the discrepancy in cold and mixing-layer gas mass only occurs when the structure is filamentary, which {induces more cooling} from the intracluster medium in Perseus simulations. In M87 simulations where cold gas is concentrated in the core, it is easily destroyed by AGN outbursts. To further demonstrate the overcooling effect induced by a mixing layer that is not resolved, we perform a one-dimensional test study of a $T\sim10^4\,{\rm K}$ infinite gas sheet in pressure equilibrium with a 1\,keV fully-ionized plasma (electron density $n_e=0.01\,{\rm cm}^{-3}$, and metallicity $Z=0.02$). The temperature distribution along the 1D tube, perpendicular to the cold gas sheet, follows:
\begin{equation}
k_{\rm B}T=
\begin{cases}
10^{-3}\,{\rm keV},\ &|x|\leq dx_0,\\
1\,{\rm keV}-0.999\, e^{-\frac{(|x|-dx_0)^2}{2\times (0.2kpc)^2}}\,{\rm keV},\ &|x|>dx_0,
\end{cases}
\end{equation}
where $dx_0=0.244\,{\rm kpc}$ is equivalent to the smallest resolution element in P8.7hr. The test tube is set up between $x=\pm8\,{\rm kpc}$ with an initial resolution of $dx_0$, and is then refined 4 times in four additional simulations, each time reducing the resolution by a factor of 2, i.e., $dx_i=2^{-i}\,dx_0$. {The outer boundaries at both ends of the tube are setup to allow the gas to freely flow in and out of the domain (i.e., the ``outflow'' boundary condition in \texttt{Enzo}).} The gas is allowed to evolve passively for 200\,Myr. For illustrative purposes, we shut off radiative cooling below $T=8,000\,{\rm K}$ to compare the growth rate of neutral hydrogen onto the cold gas sheet, and note that cooling below the threshold requires more aggressive refinement to resolve. This does not however stop adiabatic cooling within the cold gas sheet in high resolution simulations. 

Panels in Figure~\ref{fig:mix} show the evolution of neutral hydrogen mass, and the final density and temperature distributions of the five simulation tests performed at different resolutions. While there is a similar transition from cold dense gas within 100\,pc, to hot dilute plasma beyond 600\,pc, the transition location changes significantly with resolution. Simulation $dx_0$ requires a 244\,pc-mixing layer to account for the transition from ionized to neutral gas, while $dx_4$ does the same with 15\,pc. Meanwhile, because in $dx_0$ the cold gas sheet is only occupying 1 cell width, it cannot achieve pressure equilibrium with the ambient hot plasma. The unresolved mixing layer, as well as the unbalanced pressure in coarse resolution simulations leads to a higher growth rate of neutral hydrogen, which is illustrated in the third panel. Because the coarse grid sampling of the temperature profile leads to different initial neutral gas masses ($M_{\rm HI}$) compared to fine grids, in this plot we show the change of $M_{\rm HI}$, normalized to a surface area of $1\,{\rm kpc}^2$. In all 5 simulations, the growth rate of neutral hydrogen in the first 50\,Myr is similar. Beyond 50\,Myr, $dx_0$ continues to grow at a rate of $\sim10^8\,M_\sun\,{\rm Gyr}^{-1}\,{\rm kpc}^{-2}$, while $dx_{2,3,4}$ have settled to a lower rate $\lesssim10^7\,M_\sun\,{\rm Gyr}^{-1}\,{\rm kpc}^{-2}$, an order of magnitude lower. Considering the exaggerated surface area of the mixing layers in coarse resolution simulations, this alone contributes $\sim10^{11}\,M_\sun\,{\rm Gyr}^{-1}$ more cooling once filaments form in Perseus simulations, comparable to the radiative cooling rate of the intracluster medium. To avoid the `mixing' instability, the timescale for the cold gas to achieve pressure equilibrium should be shorter than $t_{\rm cool}$ of the fastest cooling layer, namely:
\begin{equation}
\begin{split}
t_{c_{\rm s}} &= \frac{dx}{\sqrt{\frac{\gamma k_{\rm B}T_{\rm cold}}{\mu m_p}}}\lesssim t_{\rm cool} = \frac{n k_{\rm B} T_{\rm mix}}{(\gamma-1)n^2\Lambda(T_{\rm mix})},\ {\rm therefore,}\\
dx &\lesssim 10 \left(\frac{T_{\rm cold}}{10^4\,{\rm K}}\right)^{0.5} \left(\frac{T_{\rm mix}}{10^6\,{\rm K}}\right) \left(\frac{n}{0.1\,{\rm cm}^{-3}}\right)^{-1} \left(\frac{\Lambda(T_{\rm mix})}{10^{-22}\,{\rm erg\,s}^{-1}\,{\rm cm}^{3}}\right)^{-1} \,{\rm pc}.
\end{split}
\end{equation}
Resolution on the order of $\sim10\,{\rm pc}$ is needed to resolve the mixing between the hot plasma and $10^4\,{\rm K}$ gas. A lower temperature floor will require a more aggressive refinement beyond current computational capacity for galaxy-scale simulations. We therefore conclude that the cold and mixing-layer gas mass is in overproduction due to resolution limitations. This affects ours and many other galaxy-scale simulations that adopt $\gtrsim0.1\,{\rm kpc}$ resolution. {We also note that our 1D tube test does not consider the velocity shear between the hot and cold gas, which may lead to the development of the Kelvin-Helmholtz instability that further mixes the two gas phases and modify the amount of cold gas \citep[e.g., studies of the ``cloud-crushing'' problem,][]{Armillotta2017, Sparre2020, Tan2021}. We further explore the radiatively cooling outflow properties in a companion work with $\sim30\,{\rm pc}$ resolution in \citet{Qiu2021b}.} 

While the cooling rate may be overestimated after cold filaments form, evolution of cold gas velocity is mainly governed by gravity, which can be extracted for comparison with observations. In the second to fourth panels of Figure~\ref{fig:vr}, we investigate the radial velocity distributions in each radial shell, averaged over 100 snapshots between $1-2\,{\rm Gyr}$. To facilitate comparison with known velocity observations, we plot the `molecular', `ionized', and `hot' components, and note that other components of the mixing layer, i.e., `cold' and `warm', share similar distribution with the `ionized' component. The radial velocity of the `molecular' component in P8.7hr is mostly contained within $\pm 250 \,{\rm km\,s}^{-1}$. This is consistent with ALMA observations of the molecular gas in central cluster galaxies, which also find a low line-of-sight velocity up to a few$\times 100 \,{\rm km\,s}^{-1}$ \citep{Russell2019}. The `ionized' component has $|v_r|\lesssim500-1,000\,{\rm km\,s}^{-1}$, which is analogous to the velocity distribution of the H$\alpha$ filaments with line widths $<600\,{\rm km\,s}^{-1}$ \citep{McDonald2012}. Lastly, the `hot' component of the simulated gas is mostly comprised of the weakly disturbed intracluster medium. While there are high velocity components on the order of $\sim1,000\,{\rm km\,s}^{-1}$ in P8.7hr, their mass is significantly lower, making them difficult to detect without spatially and kinetically resolved X-ray data. Overall, all three velocity measurements of the simulated multiphase gas appear consistent with observational constrains. }

\section{AGN Duty Cycle in Simulated M87}\label{app:duty}
{
In Section~\ref{sec:AGN} we show that the simulated AGN evolutions display cyclic behaviors on timescales of $\sim100\,{\rm Myr}$ in M8.3, M8.7, and M9.0. In this section we further analyze the duty cycle of the simulated AGNs in order to understand the fraction of time AGNs are active in elliptical galaxies like M87. In Figure~\ref{fig:duty} we show the probability distribution of AGN jet luminosity ($L_{\rm J}$) in the standard M87 simulations featuring 3 different mass-loading factors. All three plots reveal a similar bimodal distribution with 2 peaks -- one low luminosity peak $\lesssim10^{43}\,{\rm erg\,s}^{-1}$, and another high luminosity peak between $3-6\times10^{43}\,{\rm erg\,s}^{-1}$. Both peaks shift to lower $L_{\rm J}$ in simulations with higher $T_{\rm out}$. If we separate the bimodal distribution by the bin with the lowest probability between the peaks, which are respectively $L_{\rm tran} =3,\,2,\,1.26\times10^{43}\,{\rm erg\,s}^{-1}$ in M8.3, M8.7, and M9.0, we can define the AGN duty cycle ($f_{\rm duty}$) as evolution stages when $L_{\rm J} > L_{\rm tran}$. This results in a duty cycle $f_{\rm duty}=0.24,\,0.35,\, 0.51$ with increasing $T_{\rm out}$. Altogether, the opposite trends in peak jet luminosity and duty cycle as a function of $T_{\rm out}$ yield an average AGN luminosity that is similar among the three simulations, as shown in Table~\ref{tab:params}. This also means that jet-mode feedback is operating between $24-51\%$ of the time for the AGNs in the simulated M87 galaxy.

\begin{figure*}[t!]
\centering
\includegraphics[width=180mm]{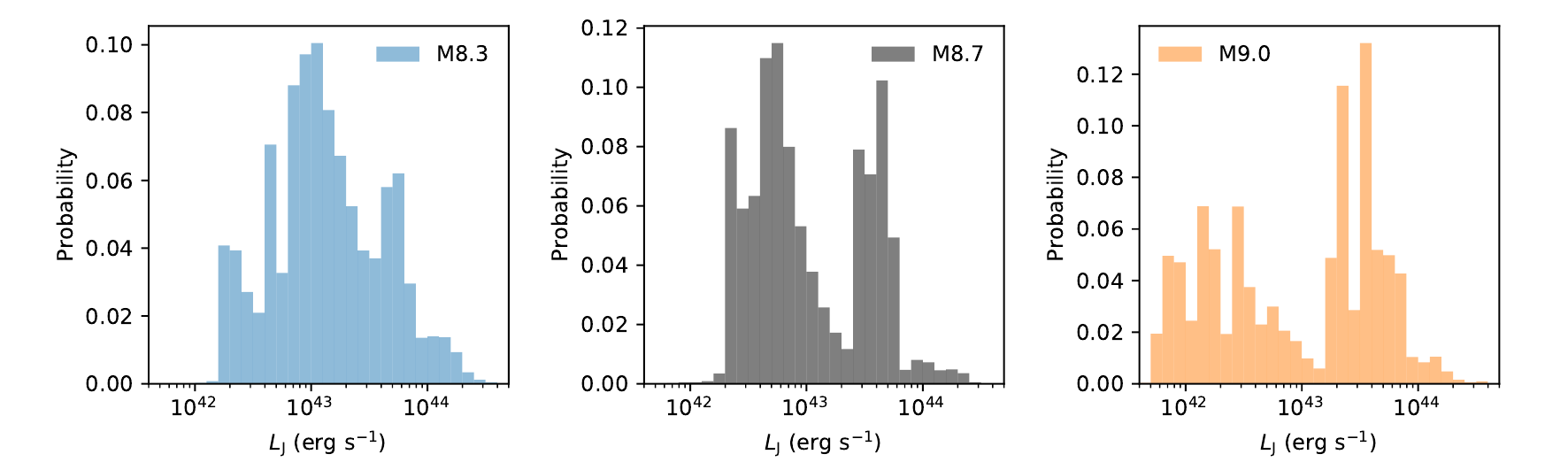}
\caption{The probability distribution of AGN jet luminosity in simulations M8.3, M8.7, and M9.0 over 2\,Gyr.}
\label{fig:duty}
\end{figure*}

For completeness, we also consider the duty cycle for the emission of radiation from AGNs in giant elliptical galaxies. Due to the low radiative efficiency of SMBHs accreting at low Eddington ratio, the radiation emitted by the AGN may not be luminous enough to be detected over a significant amount of their lifetime. Assuming a radiative efficiency ($\varepsilon_{\rm rad}$) function laid out in \citet{Inayoshi2019}:
\begin{equation}
\log{\varepsilon_{\rm rad}} = 
\begin{cases}
 -1-(0.0162/\dot{m})^4,\ &{\rm for}\ 0.023\leq\dot{m}, \\
 -0.807+0.27\log{\dot{m}},\ &{\rm for}\ 10^{-4}<\dot{m}<0.023, \\
 -1.749-0.267\log{\dot{m}}-0.07492\left(\log{\dot{m}}\right)^2,\ &{\rm for}\ 10^{-8}<\dot{m}\leq 10^{-4},
\end{cases}
\end{equation}
where $\dot{m}=\dot{M}_{\rm BH}/\dot{M}_{\rm Edd}$ is the Eddington accretion ratio. For $M_{\rm BH}=3.5\times10^{9}\,M_\sun$ in M87, $\dot{M}_{\rm Edd}\approx78\,(\eta/0.1)^{-1}\,M_\sun\,{\rm yr}^{-1}$. The fraction of time each simulated AGN spends at radiative luminosity $L_{\rm R}=\varepsilon_{\rm rad}\,\dot{M}_{\rm BH} c^2>10^{43}\,{\rm erg\,s}^{-1}$ is therefore $f_{\rm R43}=$ 6.6\%, 3.2\%, 5.5\% in simulations M8.3, M8.7, M9.0, respectively. This is roughly an order of magnitude lower than the jet duty cycle, which corroborates the expectation that jet-mode is the main feedback mechanism in giant elliptical galaxies.
}

\section{Density, Temperature and Early $t_{\rm cool}$ Profiles}\label{app:profiles}
{In the main text we examine $t_{\rm cool}$($/t_{\rm ff}$) profiles of the simulated clusters in the later half of their evolution ($t>1\,{\rm Gyr}$) to separate different jet outflow models and assess their ability in regulating the ICM properties. For completeness, here we plot $t_{\rm cool}$ profiles in the earlier half ($t<1\,{\rm Gyr}$), as well as the electron number density and temperature profiles to compare with observations of Perseus and M87. 

\begin{figure*}[b!]
\centering
\includegraphics[width=180mm]{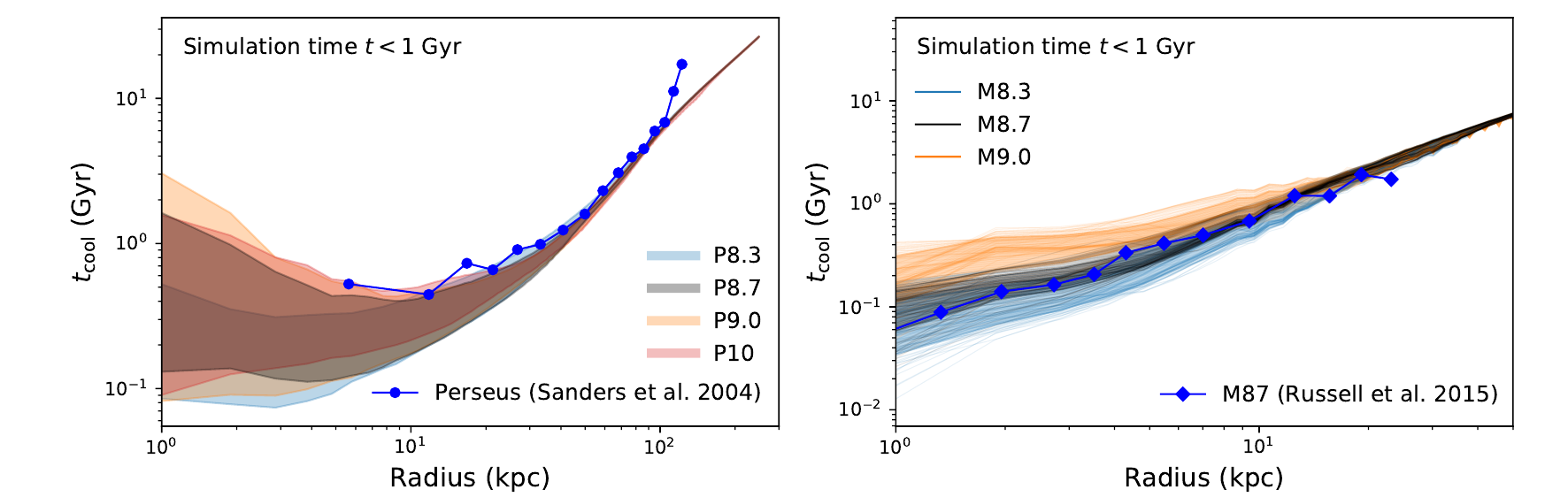}
\caption{$t_{\rm cool}$ profiles in the early stages of the simulated clusters. Thin solid lines are taken from 100 snapshots at $t<1\,{\rm Gyr}$. The widths of the colored bands represent the $\pm \sigma$-span from the mean profile in each simulation in log scale. Due to the strong overlap of the profiles in Perseus simulations, individual profiles are omitted in the left panel.}
\label{fig:earlytcool}
\end{figure*}

As noted in Section~\ref{sec:AGN}, the AGN in Perseus simulations only start to operate after $t\approx0.3\,{\rm Gyr}$. The radial profiles of $t_{\rm cool}$, $n_e$, and $T$ in the first half of the Perseus simulations are therefore similar among different outflow models (Figs.~\ref{fig:earlytcool} and \ref{fig:Perseus}). Due to overcooling, $t_{\rm cool}$ and $T$ profiles at $t<1\,{\rm Gyr}$ are lower compared with observational data in \citet{Sanders2004}. $n_e$ profiles, on the other hand, are all in general agreement with observations, because not much of the plasma has cooled and formed filaments. In the second half of the simulations ($t>1\,{\rm Gyr}$), when the outflows have had enough time to heat and regulate the thermal state of the ICM, $t_{\rm cool}$ and $T$ profiles beyond $r\approx5\,{\rm kpc}$ rise to higher values, with runs P8.7 and P9.0 most in line with observational data (Figs.~\ref{fig:profiles} and \ref{fig:Perseus}). Due to the continuous overproduction of cold gas in the simulations, however, $n_e$ between $10-100$\,kpc in the second half of the simulations is reduced by a small factor $<2$ in P8.7 and P9.0 compared with observational data. 

Plasma properties of the central $5\,{\rm kpc}$ in the Perseus cluster, we note, are hard to constrain observationally due to the unknown composition of the X-ray cavities. In our simulated Perseus cluster, the central region contains extremely hot components that lead to average temperatures up to $\sim10^9\,{\rm K}$, as shown in Fig.~\ref{fig:Perseus}. Observationally, bremsstrahlung emission from such hot, dilute plasma is much fainter compared with the surrounding $\sim10^7\,{\rm K}$ plasma, making it difficult to detect \citep[see e.g., studies that use Sunyaev-Zel'dovich effect to constrain the cavity composition, such as][]{Pfrommer2005, Colafrancesco2005, Abdulla2019, Ehlert2019, Marchegiani2021}.

\begin{figure*}[h!]
\centering
\includegraphics[width=180mm]{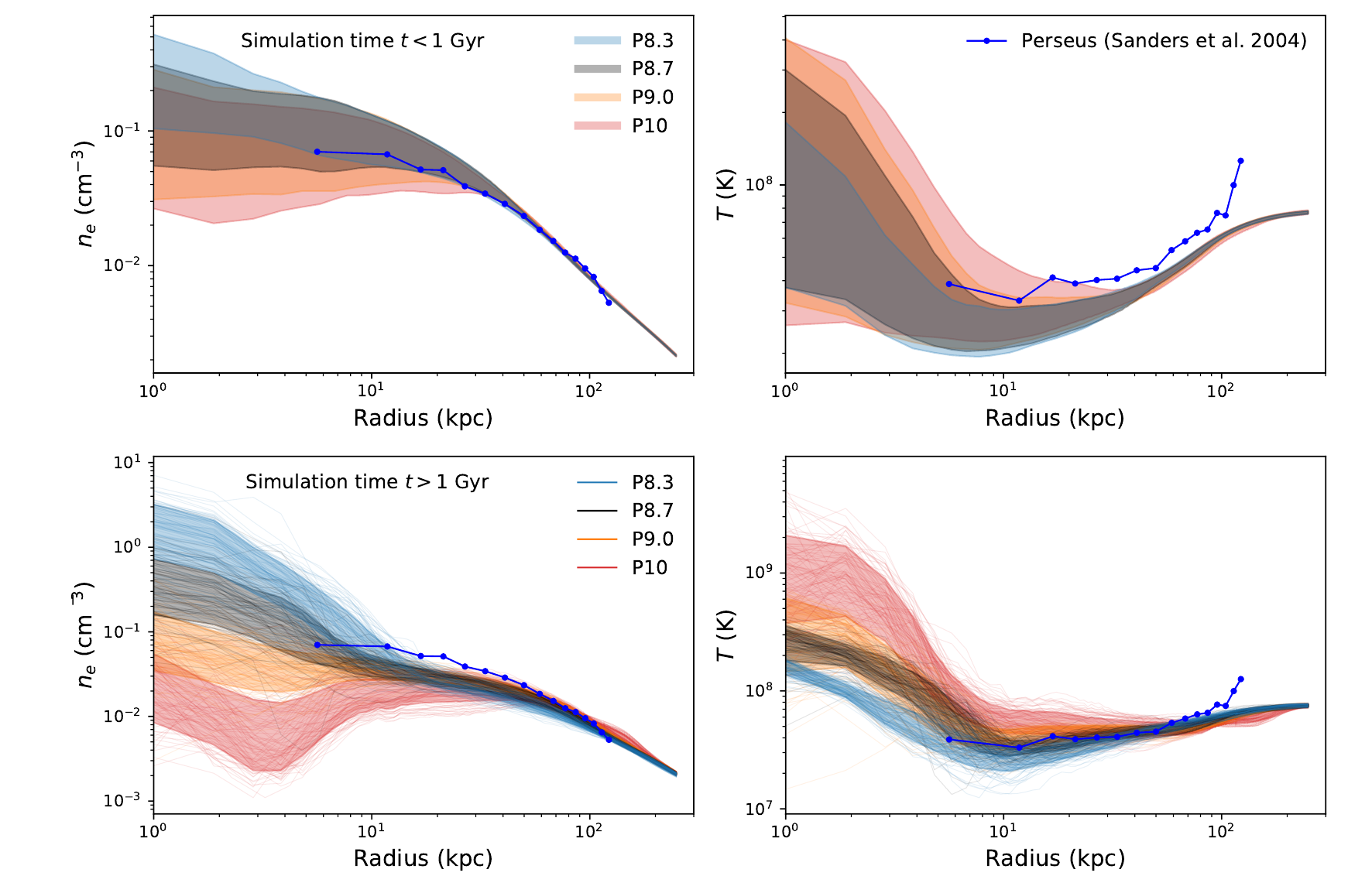}
\caption{Volume-weighted electron number density (left column) and mass-weighted temperature (right column) profiles of the hot plasma ($k_{\rm B} T>0.1\,{\rm keV}$) in Perseus simulations. Top panels show the profiles in the first half ($t<1\,{\rm Gyr}$) of the simulations, while the bottom panels show the second half. Lines and bands have same meaning as in Fig.~\ref{fig:earlytcool}. Similar to Fig.~\ref{fig:earlytcool}, individual profiles in the top panels are omitted.}
\label{fig:Perseus}
\end{figure*}

In M87 simulations, due to the short AGN cycles, $t_{\rm cool}$ profiles for both halves (before and after 1\,Gyr) of the run M8.7 are in agreement with observation data (Figs.~\ref{fig:profiles} and \ref{fig:earlytcool}). The agreement is also seen in the $n_e$ and $T$ profiles for $t<1\,{\rm Gyr}$ in simulations M8.7 and M9.0 (Fig.~\ref{fig:M87}). Note that in the central 2\,kpc of M87, two temperature components have been identified observationally \citep{Russell2015}. Here we compare with the volume-filling, 2-keV component. After 1\,Gyr, however, the temperature of the ICM in the outer region reduces below $2\times10^7\,{\rm K}$, due to the long term radiative cooling of the plasma. This also results in the elevated $n_e$ profile around 7\,kpc by a factor $\approx2$ compared with observational data. Overall, comparing the simulated ICM with observations, the AGN-driven outflows modeled with $T_{\rm out}=10^{8.67}-10^{9}\,{\rm K}$ results in $t_{\rm cool}$, $n_e$, and $T$ profiles roughly consistent with X-ray data (variations within a factor of 2). The main source of variation comes from ({\it i}) the overproduction of cold gas which reduces plasma density, and ({\it ii}) the idealized modeling of isolated clusters which cannot capture the cosmic evolution of the outer regions. 

\begin{figure*}[h!]
\centering
\includegraphics[width=180mm]{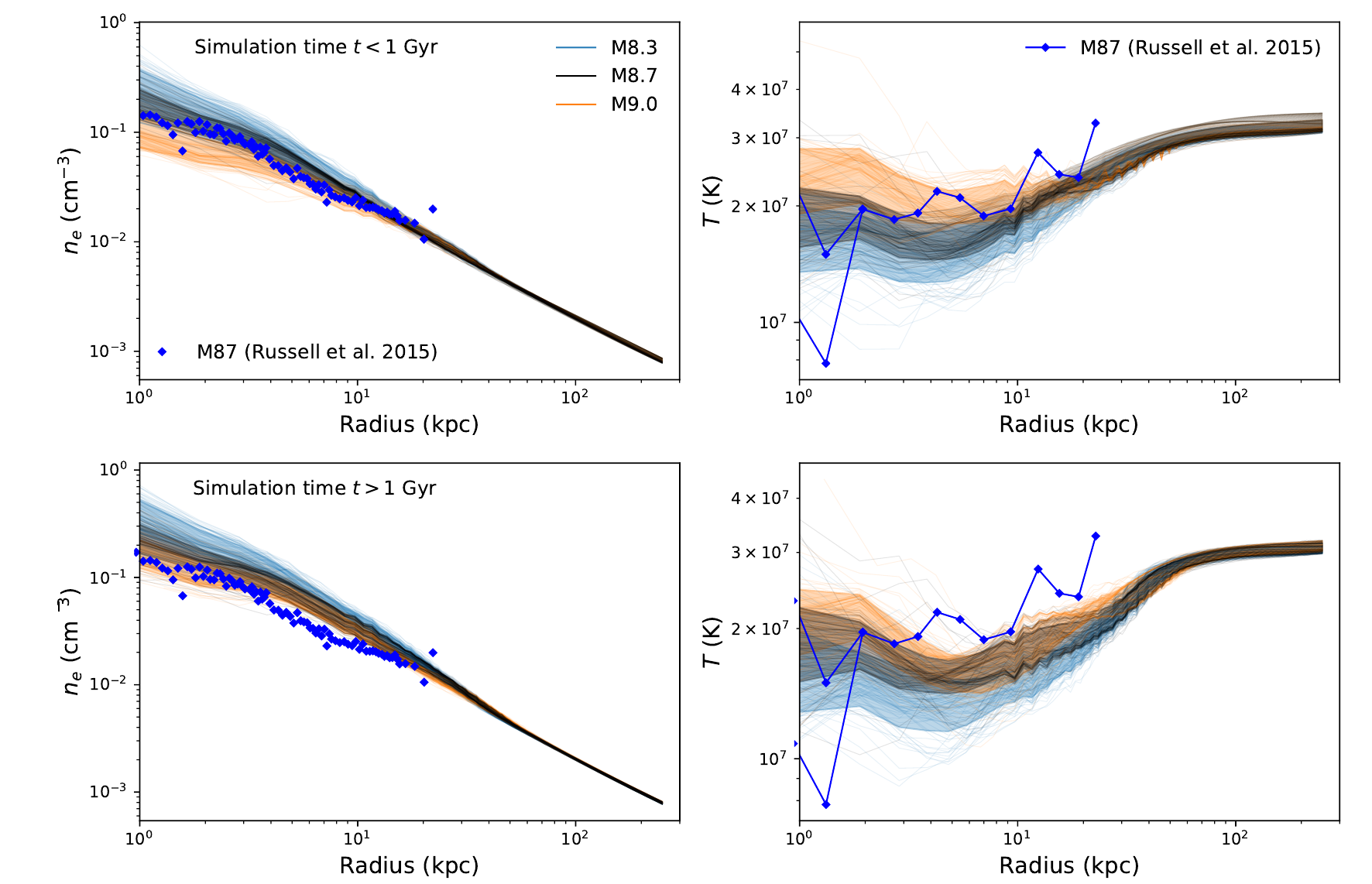}
\caption{Volume-weighted electron number density (left column) and mass-weighted temperature (right column) profiles of the hot plasma ($k_{\rm B} T>0.1\,{\rm keV}$) in M87 simulations. Top panels show the profiles in the first half ($t<1\,{\rm Gyr}$) of the simulations, while the bottom panels show the second half. Lines and bands have same meaning as in Fig.~\ref{fig:earlytcool}.}
\label{fig:M87}
\end{figure*}

}

\bibliography{reference}{}
\bibliographystyle{aasjournal}

\end{CJK*}

\end{document}